\documentclass[prd,reprint,nofootinbib,superscriptaddress,preprintnumbers,longbibliography,10pt,toc]{revtex4-2}
\usepackage{graphicx,epsfig,psfrag,bm,amsmath,amssymb}
\usepackage{dcolumn}
\usepackage{bm}
\usepackage[dvipsnames]{xcolor}
\usepackage{braket}
\usepackage{mathrsfs,amsfonts,color}
\usepackage{comment}
\usepackage{soul}
\usepackage[caption=false]{subfig}
\usepackage{setspace}
\usepackage{siunitx}
\usepackage[page,toc,titletoc,title]{appendix}
\usepackage{hyperref}
\usepackage[capitalise]{cleveref}

\definecolor{myred}{rgb}{0.7, 0, 0}
\definecolor{myblue}{rgb}{0, 0, 0.7}
\definecolor{mygreen}{rgb}{0.04, 0.7, 0.5}
\definecolor{mygray}{rgb}{0.1, 0.1, 0.1}

\hypersetup{colorlinks,citecolor=myred,linkcolor=myblue,urlcolor=myblue,linktoc=all}

\DeclareRobustCommand{\Eq}[1]{Eq.~(\ref{#1})}

\newcommand{\Fig}[1]{Fig.~\ref{fig:#1}}

\newcommand{\be}{\begin{eqnarray}}
\newcommand{\ee}{\end{eqnarray}}

\newcommand{\Mpl}{M_{\rm Pl}}

\newcommand{\x}{\mathbf{x}}

\newcommand{\lt}{\ell_+}

\newcommand{\Dl}{\Delta_\ell}

\newcommand{\w}{\omega}

\newcommand{\grad}{\nabla}

\newcommand{\rhoDM}{\rho_{\text{DM}}}
\renewcommand{\vec}[1]{\boldsymbol{#1}}

\newcommand{\beq}{\begin{equation}} 
\newcommand{\eeq}{\end{equation}} 
\newcommand{\bead}{\begin{aligned}}  
\newcommand{\eead}{\end{aligned}} 
\renewcommand{\[}{\left[}
\renewcommand{\]}{\right]}
\renewcommand{\(}{\left(}
\renewcommand{\)}{\right)}
\newcommand{\leri}[1]{\left(#1 \right)}
\newcommand{\cmt}[1]{}
\def\bes{\begin{equation*}}
\def\ees{\end{equation*}}
\def\bmat{\left(\begin{matrix}}
\def\emat{\end{matrix}\right)}

\newcommand{\kd}{q}
\newcommand{\ma}{m_a}
\newcommand{\kres}{k_{\rm res}}
\newcommand{\dm}{{\delta m}}
\newcommand{\dk}{{\delta k}}
\newcommand{\xd}{x_{\delta}}
\newcommand{\lw}{{l}_{\rm wire}}
\newcommand{\kvir}{{k}_{\rm vir}}

\newcommand{\vect}[1]{\boldsymbol{#1}}

\newcommand{\kdres}{q_{\rm res}}
\newcommand{\abs}[1]{\left| #1 \right|}

\usepackage[nolist]{acronym}

\def\UMD{\small{Maryland Center for Fundamental Physics, University of Maryland, College Park, MD 20742, USA}}

\usepackage{tikz,xcolor,hyperref}
\definecolor{lime}{HTML}{A6CE39}
\DeclareRobustCommand{\orcidicon}{
	\begin{tikzpicture}
	\draw[lime, fill=lime] (0,0) 
	circle [radius=0.16] 
	node[white] {{\fontfamily{qag}\selectfont \tiny ID}};	\draw[white, fill=white] (-0.0625,0.095) 
	circle [radius=0.007];	\end{tikzpicture}
	\hspace{-2mm}}
\foreach \x in {A,...,Z}{
	\expandafter\xdef\csname orcid\x\endcsname{\noexpand\href{https://orcid.org/\csname orcidauthor\x\endcsname}{\noexpand\orcidicon}}
}

\date{\today}

\begin{document}

\interfootnotelinepenalty=10000 

\title{Momentum and Matter Matter for Axion Dark Matter Matters on Earth}

\author{Abhishek Banerjee\orcidA{}\,}
\affiliation{\UMD}
\author{Itay M. Bloch\orcidB{}\,}
\affiliation{Theoretical Physics Group, Lawrence Berkeley National Laboratory, Berkeley, CA 94720, U.S.A.}
\affiliation{Berkeley Center for Theoretical Physics, University of California, Berkeley, CA 94720, U.S.A.}
\author{Quentin Bonnefoy\orcidC{}\,}
\affiliation{Universit\'e de Strasbourg, CNRS, IPHC UMR7178, 23 rue du Loess, 67037 Strasbourg, France}
\author{Sebastian~A.~R.~Ellis\orcidD{}\,}
\affiliation{D\'epartement de Physique Th\'eorique, Universit\'e de Gen\`eve, 
24 quai Ernest Ansermet, 1211 Gen\`eve 4, Switzerland}
\author{Gilad Perez\orcidE{}\,}
\affiliation{Department of Particle Physics and Astrophysics, Weizmann Institute of Science, 234 Herzl Street, Rehovot 7610001, Israel}
\author{Inbar Savoray\orcidF{}\,}
\affiliation{Theoretical Physics Group, Lawrence Berkeley National Laboratory, Berkeley, CA 94720, U.S.A.}
\affiliation{Berkeley Center for Theoretical Physics, University of California, Berkeley, CA 94720, U.S.A.}
\author{Konstantin Springmann\orcidG{}\,}
\affiliation{Department of Particle Physics and Astrophysics, Weizmann Institute of Science, 234 Herzl Street, Rehovot 7610001, Israel}
\author{Yevgeny V.~Stadnik\orcidH{}\,}
\affiliation{
School of Physics, The University of Sydney, Sydney, NSW 2006, Australia}

\begin{abstract}
    We investigate the implications of matter effects to searches for axion Dark Matter on Earth. 
    The finite momentum of axion Dark Matter is crucial to elucidating the effects of Earth on both the axion Dark Matter field value and its gradient. 
    We find that experiments targeting axion couplings compatible with canonical solutions of the strong CP puzzle are \emph{likely not affected} by Earth's matter effects. However, experiments sensitive to lighter axions with stronger couplings can be significantly affected, with a significant part of the parameter space suffering from a \emph{reduced axion field value}, and therefore decreased experimental sensitivity. In contrast, the spatial gradient of the axion field can be enhanced along Earth's radial direction, with important implications for ongoing and planned experiments searching for axion Dark Matter.
\end{abstract}

\maketitle

\onecolumngrid
\tableofcontents

\begin{acronym}
    \acro{EOM}{Equation of Motion}
    \acrodefplural{EOM}[EOMs]{Equations of Motion}
    \acro{DM}{Dark Matter}
    \acro{SM}{Standard Model}
    \acro{BSM}{Beyond the Standard Model}
    \acro{ULDM}{Ultralight Dark Matter}
    \acro{EM}{Electromagnetic}
    \acro{KG}{Klein-Gordon}
\end{acronym}

\section{Introduction}

Light pseudoscalar fields are well-motivated candidates for new physics \ac{BSM} and are the target of an extensive experimental program. 
One example of such a pseudoscalar is the axion, which can play a major role in the solution to the strong CP problem~\cite{Peccei:1977hh,Peccei:1977ur,Weinberg:1977ma,Wilczek:1977pj}. Other light pseudoscalars may take part in alternative solutions to the strong CP problem~\cite{Dine:2024bxv}, as well as in solutions to other puzzles and problems of the \ac{SM} -- such as the flavor puzzle~\cite{Froggatt:1978nt,Gelmini:1982zz,Anselm:1985bp,Feng:1997tn,Calibbi:2016hwq} and the hierarchy problem~\cite{Graham:2015cka,Banerjee:2018xmn,Chatrchyan:2022dpy}. Light axions with masses below an $\text{eV}$ have garnered further interest as leading \ac{DM} candidates~\cite{Preskill:1982cy,Abbott:1982af,Dine:1982ah}.
Since the 1980s, a significant experimental effort has targeted the particularly interesting regime where the axions are ultralight fields that solve the strong CP puzzle \emph{and} constitute the \ac{DM} of the universe (see, e.g., Refs.~\cite{DiLuzio:2020wdo,Adams:2022pbo,OHare:2024nmr} for recent reviews). Indeed, many of these experiments rely on the axion field value fitted from the local \ac{DM} abundance, 
i.e.,
$m_a^2 \langle a^2(t) \rangle_t = \rhoDM$,
where $\rhoDM \sim 0.4~\text{GeV/cm}^3$ is the inferred local \ac{DM} density~\cite{deSalas:2020hbh}, and $m_a$ is the axion mass.

The relic energy density of light, wave-like \ac{DM} arises due to the oscillations of the field in its potential. The value of $\langle a(t)^2\rangle $ computed above, corresponds to the average axion field in vacuum. However, in an environment of large \ac{SM} energy density, such as a star or planet, the amplitude of $a$ could drastically change. For example, axions can exhibit superradiance when scattering off of black holes~\cite{Mehta:2020kwu,Unal:2020jiy,Hoof:2024quk,Witte:2024drg}. Similarly, neutron stars have been shown to have a significant impact on the axion field value~\cite{Hook:2017psm,Balkin:2020dsr,Prabhu:2021zve,Budnik:2020nwz,Balkin:2021wea,Balkin:2021zfd,Noordhuis:2023wid,Balkin:2023xtr,Gomez-Banon:2024oux,Kumamoto:2024wjd} through their extreme density of both matter and electromagnetic energy. Furthermore, white dwarf stars have been shown to also have an important impact on the axion field value~\cite{Balkin:2022qer}. 
The backreaction of the modified axion field value on these stars has been studied in \cite{Balkin:2022qer,Balkin:2023xtr,Gomez-Banon:2024oux,Kumamoto:2024wjd}, and was used to probe various axion models. Importantly, these last works do not rely on the axion constituting the \ac{DM}.

Recently, it was also realized that QCD-axion-like \ac{DM} models can be probed through their quadratic interactions with the \ac{SM} using clocks~\cite{Kim:2022ype}. This is particularly interesting given the recent progress related to laser excitations of the Th-229 nucleus~\cite{Tiedau:2024obk,Zhang:2024ngu}, which probably already exceeds laboratory-based bounds of the models mentioned above~\cite{Caputo:2024doz,Fuchs:2024edo}. 
However, the \ac{DM} solution to the axion's \acp{EOM} can also be significantly altered via these quadratic interactions in the presence of a matter source. 
The impact of quadratic couplings to matter effects for ultralight \ac{DM} was studied initially in~\cite{Hees:2018fpg} for scalar (CP-even) fields, and the implications for model building were studied in~\cite{Banerjee:2022sqg}. In these works, it was shown that under the assumption of a homogeneous temporally-oscillating solution at infinity, required for the field to play the role of \ac{DM}, the field value on the surface of Earth can be strongly screened or strongly enhanced (depending on the sign of the quadratic coupling) for some choices of the quadratic coupling. Recently, a similar analysis was carried out for QCD-axion-like ultralight \ac{DM} in Refs.~\cite{Bauer:2024yow,Bauer:2024hfv}. Like in the scalar case analyses in Refs.~\cite{Hees:2018fpg,Banerjee:2022sqg}, these recent analyses took the zero-momentum limit for the axion \ac{DM} field, and found an apparent divergence of the axion field amplitude near Earth when the axion decay constant approaches a set of discrete values.

In this paper, we carefully analyze how Earth affects wave-like \ac{DM} with attractive quadratic couplings, going beyond the zero-momentum limit (for such studies on scalar-field quadratic interactions with the opposite sign, see Refs.~\cite{Stadnik:2024highlighting,Day:2024blowing,Tilburg:2024wake}).
The axion's classical \acp{EOM} with the boundary condition set by the \ac{DM} solution at infinity lead to a Schr\"odinger eigenvalue problem for the scattering of a wave off a potential well. Hence, the system exhibits quantum-like phenomena wherein the spectrum admits a discrete set of bound states.
Furthermore, discrete values of the input parameters exist such that zero-energy bound states can be found. In these cases, one can excite a bound state by sending states with zero momentum at infinity, leading to a ``runaway-behavior". 
This explains the findings of Refs.~\cite{Bauer:2024yow,Bauer:2024hfv}, where the axion momentum was set to zero in Earth's reference frame. 
As Earth is not static in the Solar frame, and the Sun is not static in either the Galactic frame or that of the \ac{DM} (see, however, Refs.~\cite{Banerjee:2019epw,Banerjee:2019xuy,Budker:2023sex}), one needs to reconsider the claimed effect once the \ac{DM}'s gradient energy is added to the problem. We show that the divergent solution found in Refs.~\cite{Bauer:2024yow,Bauer:2024hfv} becomes a rather exotic possibility, and typical \ac{DM} momentum distributions such as a boosted Maxwell-Boltzmann distribution (or any 3D distribution which does not diverge for $k \to 0$) exhibit no such behaviour. Nevertheless, we find that there are several profound implications for the \ac{DM} dynamics and the corresponding experimental signatures. 
Specifically, the finite momentum of the axion field can lead to 
resonances in the axion field value for certain values of the decay constant $f_a$. 
The precise conditions required for the existence of such resonances will be discussed. 
Importantly, off-resonance, the axion amplitude and axion gradient are also dramatically affected; while the axion amplitude is often \emph{suppressed}, resulting in possibly reduced sensitivity for certain experiments (see, e.g.,~\cite{Schulthess:2022pbp,Abel:2017rtm,Roussy:2020ily,Madge:2024aot,Fuchs:2024edo,JEDI:2022hxa,Zhang:2022ewz,Fox:2023xgx,Blum:2014vsa,Caloni:2022uya,Badziak:2024szg,Zhang:2023lem,Alda:2024xxa}), the axion gradient is typically \emph{enhanced}, possibly improving the sensitivity of spin-magnetic (axion) experiments (see, e.g.,~\cite{Flambaum_Patras_2013,Stadnik:2013raa,Stadnik_2017_book-thesis,Abel:2017rtm,Wu:2019exd,Garcon:2019inh,smorra_direct_2019,Bloch:2019lcy,Bloch:2021vnn,Jiang:2021dby,Bloch:2022kjm,Lee:2022vvb,Abel:2022vfg,Wei:2023rzs,Xu:2023vfn,Gavilan-Martin:2024nlo,Fierlinger:2024aik,JEDI:2022hxa}). 

The paper is structured as follows. In Section~\ref{sec:overview}, we present an overview of our results, categorizing how different axion parameters can lead to different modifications due to Earth. 
In Section~\ref{sec:mathbasics}, we solve the axion profile for a field with a well-defined momentum far from Earth, showing how both the field and its gradient are modified for a few sampled parameters. 
In Section~\ref{sec:lowL}, we describe the phenomenology arising from the first two terms in the multipole expansion of the axion field (the monopole and dipole). 
These terms dominate the axion behavior when its virial momentum is {much} smaller than the inverse radius of Earth. 
While most of this work focuses on the deformations of the axion field at the surface of Earth, within this section, we also discuss the modifications of the axion field and gradient deep within Earth, and far outside it. 
In Section~\ref{sec:ExpConseq}, we discuss how our results affect existing and proposed experiments. We finish the paper with our conclusions in Section~\ref{sec:conclusions}, commenting on the importance of this work, and emphasizing future refinements that should be done to the calculations presented here, as well as other interesting questions to explore. 
We also include several appendices within this work. In Appendix~\ref{app:virial}, we discuss how to compute the power spectral densities of stochastic quantities with a virially distributed velocity, rather than ones with a well-defined momentum. 
For completeness, we show in Appendix~\ref{app:conv} that the finite momentum limit fully regulates all divergences. In Appendix~\ref{app:resonances}, we discuss the effect of resonances, demonstrating how the virial integration regulates their heights. In Appendix~\ref{app:sommer}, we connect our resonances to the presence of near-threshold bound states. In Appendix~\ref{app:pert}, we use perturbation theory to examine when the matter effects are negligible. 
In Appendix~\ref{app:Vpt2}, we expand on some of the discussion in section~\ref{sec:ExpConseq}, further commenting on the  detectability of Earth-affected axions through their coupling to photons.

\section{Overview of our Main Results}\label{sec:overview}

We start by giving a brief overview of the results we find. 
The effect of the medium is to change the potential of the axion field, which to leading order reduces the mass of the axion. 
Inside Earth, this shift is \cite{Hook:2017psm}
\begin{align}
    {\dm}_\oplus \sim 10^{-9}\,\text{eV} \times \frac{10^9\,\text{GeV}}{f_a} \ . 
    \label{eq:deltaMEarth}
\end{align}
This effect can be equally described in the context of an effective field theory of a light spin-0 field, by appropriately choosing the sign of the quadratic interaction to reduce the mass.

Neglecting higher-order terms in the axion cosine potential, the axion field value must satisfy the \ac{KG} equation, $(\Box + m_a^2)a = 0$. 
This equation can be easily solved in Fourier space, where we find that the axion field is a sum over $k$-modes, with each mode satisfying the dispersion relation $\w_k^2 = m_a^2 + k^2$. 
For a \ac{DM} field in vacuum, the typical (virialized) value of the momentum is $\kvir\equiv m_a\,v \sim 10^{-3} m_a$\footnote{In the Standard Halo Model (SHM), ${\bf v}={\bf v}_{\rm gal}+{\bf v}_{\oplus}$, where ${\bf v}_{\oplus}$ is defined as the velocity of Earth in the Galactic frame, with $|{\bf v}_\oplus(t)|\approx 252\pm 10\,\text{km/s}$. 
${\bf v}_{\rm gal}$ is sampled from a Maxwell Boltzmann distribution with a virial velocity $v_{\rm vir} \simeq 238\,\text{km/s}$, truncated at the Galactic escape velocity of $550\, \text{km/s}$~\cite{Baxter:2021pqo}.}. 
In the interface between matter where $\dm\neq 0$ and vacuum where $\dm=0$, the axion dispersion relation must be satisfied on both sides. 
The frequency of oscillation of a given $k$-mode does not change since energy is conserved in the presence of a time-independent matter density, so the momentum of the axion field changes from one side of the interface to the other. This is precisely analogous to \ac{EM} waves at an interface between vacuum and a medium with a non-unity refractive index. We can therefore develop an intuition for our findings from this analogy.

Let us consider the case of an \ac{EM} plane wave traveling from vacuum into a non-lossy medium (i.e., the momentum is always purely real) with a refractive index $n > 1$. 
In this scenario, the wavevector changes from $k$ to $nk \geq k$. For an axion entering a medium, the wavevector goes from $k$ to $\kd\equiv\sqrt{k^2 + {\dm}^2} \geq k$, which is equivalent to an effective refractive index for the axion wave $n \sim |{\dm}|/k$ in the limit $|{\dm}| \gg k$. 
Thus, in the large $|{\dm}|$ limit we expect significant refraction and the field value at the interface to be strongly suppressed.

\begin{figure}[h]
    \centering
    \includegraphics[width=0.8\linewidth]{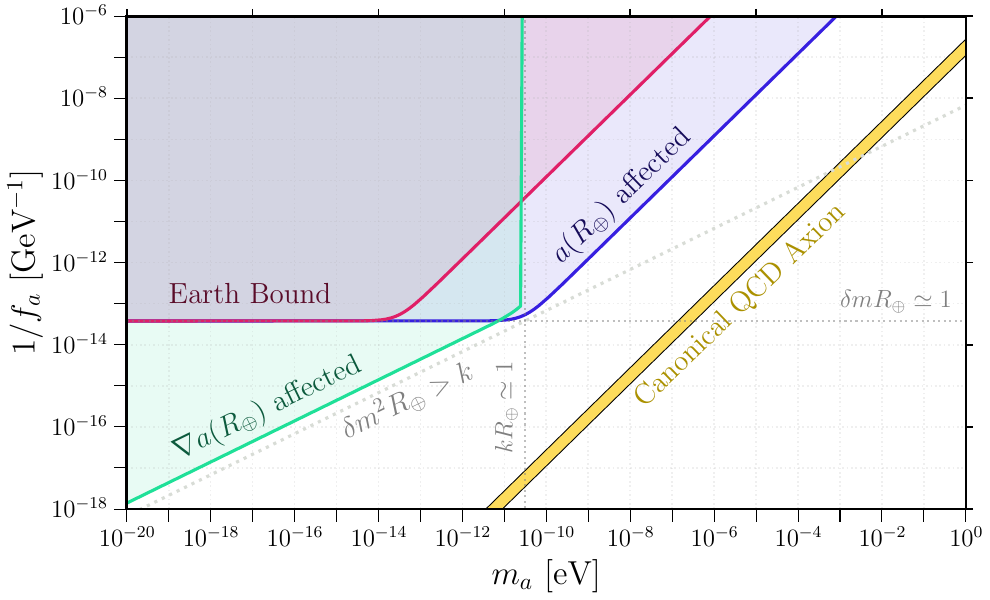}
    \caption{
    We show as blue and green shaded regions the parts of the axion parameter space impacted by our analysis. In blue, the axion field value on Earth, $a(R_\oplus)$, is affected as described in the text, with most parts having a reduced axion field value. In green, the spatial gradient of the axion field is affected. The radial gradient of the axion field is typically enhanced by a factor $1/(kR_\oplus)$, while gradients in other directions are generally suppressed. See Section~\ref{sec:lowL} for more details. In the overlap of the green and blue regions, the results of Section~\ref{sec:lowL} apply. The experimental consequences are described in Section~\ref{sec:ExpConseq}. We show as a grey dashed line the region where $\delta m^2 R_\oplus > k$, which sets the lower bound on $1/f_a$ for which Earth could have an impact on observables (including the presence of resonances, see text for further details). Using perturbation theory, we may estimate that the matter effects of Earth below this dashed line are entirely negligible. 
    It is likely that no significant change in experimental observables is present in the region between the diagonal grey dashed line and the blue/green regions, however, further work is necessary to better characterize this intermediate region. See the text for additional details. Also shown is the related bound from sourcing axions on Earth from~\cite{Hook:2017psm} (red shaded region). 
    For clarity we are not showing other bounds in this plane, such as~\cite{Schulthess:2022pbp,Abel:2017rtm,Roussy:2020ily,Madge:2024aot,Fuchs:2024edo,JEDI:2022hxa,Zhang:2022ewz,Fox:2023xgx,Blum:2014vsa,Mehta:2020kwu,Baryakhtar:2020gao,Unal:2020jiy,Hook:2017psm,Zhang_2021,Banerjee:2023bjc,Caloni:2022uya,Balkin:2022qer,Badziak:2024szg,Gomez-Banon:2024oux,Hoof:2024quk,Springmann:2024ret,Kumamoto:2024wjd,Witte:2024drg}. 
    A complete list can be found on \cite{AxionLimits}.}
    \label{fig:parspace}
\end{figure}

In this work, we solve the axion's EOM, and identify regions of the parameter space that will be significantly affected by the Earth's matter, presented in Fig.~\ref{fig:parspace}. We show that the amplitude of the axion \ac{DM} field is significantly modified (see \Fig{l3sol}) when 
\begin{equation}
    qR_\oplus\gtrsim\begin{cases}\mathcal{O}(1)\,,&kR_\oplus<1\\
    \mathcal{O}(kR_\oplus)\,,&kR_\oplus>1
    \end{cases} \ ,
\end{equation}
which corresponds to the blue region in Fig.~\ref{fig:parspace}. For the QCD axion, $q\geq k$ always.
We define canonical QCD axions, as models which obey the relationship $m_af_a\simeq m_\pi f_\pi$ (e.g., the KSVZ~\cite{Kim:1979if,Shifman:1979if} and DFSZ~\cite{Dine:1981rt,Zhitnitsky:1980tq} models, shown as a yellow band in Fig.~\ref{fig:parspace}). However, it has been shown that the relation can be modified to $m_af_a\simeq \sqrt{\epsilon}\,m_\pi f_\pi$, with $\epsilon \leq 1$ without requiring a large tuning of parameters~\cite{Hook:2018jle} (see also~\cite{DiLuzio:2021pxd,DiLuzio:2021gos}).
The condition $q>\mathcal{O}(k)$ is therefore satisfied when $\dm \gtrsim k$, corresponding to $\epsilon \lesssim 2\times 10^{-8}$.

The requirement that $q R_\oplus \gtrsim 1$ implies that the axion amplitude is only impacted for axion decay constants smaller than $f_a \lesssim 3\times 10^{13}\,\text{GeV}$. 
Together, these requirements allow us to define an axion mass-dependent critical coupling for Earth,
\begin{align}
    (f_a)^c_\oplus(m_a) \equiv \frac{3\times 10^{13}\,\text{GeV}}{\sqrt{1+v^2m_a^2 R_\oplus^2}} \ ,
    \label{eq:EarthCritFa}
\end{align}
below which Earth plays a major role in changing the field amplitude (the blue region in \Fig{parspace}).\footnote{The factor $\sqrt{1+v^2m_a^2 R_\oplus^2}$ in the denominator approximates gradient effects that arise when solving the full coupled system as calculated in \cite{Balkin:2022qer,Balkin:2023xtr}. Without taking into account gradient effects, it would be $\text{max}\left[1,m_a v R_\oplus\right]$.}
Note that for $vm_aR_\oplus\ll1$, this roughly matches the condition where sourcing is prevented by gradient energy, $R_\oplus\simeq \dm^{-1}$.

Outside Earth, the asymptotic spatial dependence of the axion field is such that $(\grad a)/a \sim k$, while inside Earth, the typical momentum scale is $q\geq k$. 
As will be discussed later, we expect that Earth's effects are in general negligible when $(\delta m R_\oplus)^2 > (k R_\oplus)$, shown as the diagonal grey dashed line in Fig.~\ref{fig:parspace}. We find that in practice, significant effects of Earth on the axion gradient only occur when $k R_\oplus < 1$ is \emph{also} satisfied.  We show this parameter space in green in Fig.~\ref{fig:parspace}, which corresponds to when the conditions
\begin{align}
    f_a \lesssim 10^{14}\,\text{GeV} \times \sqrt{\frac{10^{-12}\,\text{eV}}{m_a}}, \quad {\rm and,}\quad m_a \lesssim \frac{1}{R_\oplus v} \ ,
\end{align}
are both met. 
Importantly, in the entire green region, the radial component of $\nabla a$ is enhanced compared with the vacuum expectation of $\nabla a \sim k a_0$. This enhancement persists even in the overlap between the blue and green regions in Fig.~\ref{fig:parspace}, where the in-medium axion field value is generically \emph{damped} with respect to the local \ac{DM} field value. Conversely, the axion gradient perpendicular to Earth's surface is suppressed in the same manner as the field amplitude (and within the same blue region). 

For sufficiently small values of $\epsilon$, the Earth-induced shift in the axion mass of Eq.~\eqref{eq:deltaMEarth} leads to a negative mass-squared of the axion which can drive the axion field from its usual $a=0$ vacuum to $a=\pi f_a$\footnote{It should be understood that this is a pure vacuum effect. 
Axion particles would correspond to fluctuations $\delta a$ around the $r$-dependent vacuum expectation value, $a(r,t)=a(r)+\delta a$. 
Since strong constraints arise from changing the vacuum inside Earth, there is no point in solving such a problem.}~\cite{Hook:2017psm,Balkin:2020dsr}. This can occur for axion masses\footnote{Note that for sourcing (the formation of a non-trivial field profile) to occur, $\delta mR_\oplus>1$ must also be satisfied. Hence the object must be dense and large enough \cite{Balkin:2021zfd}.}
\begin{equation}
    m_a
    \simeq\sqrt{\epsilon}\frac{\sqrt{z}}{1+z}\frac{m_\pi f_\pi}{f_a}
    <\dm_\oplus =5.3\times10^{-11}\,\text{eV}\left(\frac{\epsilon}{8\times 10^{-15}}\right)^{1/2}\left(\frac{10^{10}\,\text{GeV}}{f_a}\right),
    \label{eq:EarthSourcing}
\end{equation}
where $z=m_u/m_d$. The region where this occurs on Earth is shown in red in Fig.~\ref{fig:parspace}. The same effect can affect the stability of white dwarfs~\cite{Balkin:2022qer}, which leads to a more stringent constraint on $\epsilon$ which is omitted from the figure for simplicity. See \Fig{GradExps} and \Fig{EfieldDCmqs} for a comparison of selected bounds. The parameter space where $8\times 10^{-15} \lesssim \epsilon \lesssim 2\times 10^{-8}$ is therefore in the regime where our analysis applies without being modified by the dynamics discussed in Refs.~\cite{Hook:2017psm,Balkin:2020dsr}. We stress that the axion's profile can be significantly affected even in regions of the parameter space where no sourcing (i.e. the axion's vacuum is $a=0$), is present due to Earth's matter effects.

As with \ac{EM} waves traversing a medium of finite extent with $n>1$, resonances may arise due to constructive interference associated with the finite size of the region $R$. 
This can also be understood considering the textbook Schr\"odinger quantum mechanics problem of a plane wave scattering off of an attractive spherical potential well, which may exhibit resonant Sommerfeld enhancement. 
The axion scatters off Earth's spherically-symmetric negative-energy step-function potential, which can lead to bound states. The finite crossing time of the axion field implies that even when the axion wavefunction has a significant overlap with a bound state, its amplitude is not divergent, but simply resonant. 
The demonstration of the relationship between the resonances and bound states can be found in Appendix~\ref{app:sommer}. 
In most of our work, we neglect the effect of these resonances. 
However, it is worth noting that for $\kvir R>1$, within the blue region, nearly any point would include the contribution from $k$ modes which are resonant, leading to non-trivial effects which are not captured by this overview. 
More broadly, resonant contributions may play a role above the diagonal grey dashed line (c.f.~Appendices~\ref{app:resonances} and~\ref{app:pert}).

\section{Solutions with Finite Asymptotic Momentum}\label{sec:mathbasics}

The origin of the matter effects lies in the non-derivative couplings of axions to nucleons. These couplings follow automatically from the defining coupling of the QCD axion to gluons (see, e.g.,~\cite{Moody:1984ba}).
Assuming the axion is responsible for the local \ac{DM} density, the amplitude of the field $a$ is sufficiently small that these effects are dominated by the quadratic coupling of axions to nucleons,
\begin{equation}
\label{eq:axion_nucleon_int}
\mathcal{L}=-\sigma_N\bar{N}N\sqrt{1-\beta\sin^2\left(\frac{a}{2f_a}\right)}\simeq \frac{1}{2}\frac{z}{(1+z)^2}\sigma_N\bar{N}N\frac{a^2}{f_a^2}\,,
\end{equation}
where $N=(p,n)^{\operatorname{T}}$ is the nucleon field, $\beta=4z/(1+z)^2$ with $z=m_u/m_d$, and $\sigma_N\simeq 50~\text{MeV}$ is the nucleon sigma term (see, e.g.,~\cite{Alarcon:2021dlz} and references therein). 
As the axion is CP odd, it is evident that in the absence of CP violation, an axion coupling to a CP-even nucleon operator would naturally arise at the quadratic order. 
This coupling corresponds to $d_{m_N}^{(2)}=-z\sigma_N\Mpl^2/[(1+z)^2f_a^2]$ in the notation of, e.g.,~\cite{Hees:2018fpg}, where $\Mpl$ is the Planck scale. 
In particular, the interaction in Eq.~(\ref{eq:axion_nucleon_int}) implies that the axion mass is \textit{reduced} in a background of finite nucleon number density 
$\braket{\bar{N}N}\simeq n\neq 0$ such as Earth. 
We can write the effective in-medium axion mass-squared as
\begin{equation}
\left(m_a^\text{eff}\right)^2=m_a^2-\dm^2=\frac{z}{(1+z)^2}\frac{1}{f_a^2}\left( \epsilon\, m_\pi^2f_\pi^2 - \sigma_{N}\,n \right),
\end{equation}
where $\dm^2>0$ is the shift in the mass-squared due to the finite density background.

The axion's EOM in three dimensions in the presence of Earth, which we model by a perfect sphere of constant density $\rho_\oplus \simeq n\,m_N \simeq 5.5\,\text{g}/\text{cm}^3$ and radius $R\simeq 6400\,\text{km}$, is now
\begin{align}
\label{eq:EOM1}
    \begin{cases}
        \(\Box+m_a^2\)a(\vect{r},t)=0 \,,&  r>R \\
        \(\Box+m_a^2-\dm^2\)a(\vect{r},t)=0 \,,& r<R
    \end{cases}\,.
\end{align}
We have introduced the quantity $r = |\vect{r}-\vect{r}_\oplus|$ to denote the distance in the radial direction in Earth-centric coordinates. We assume that the axion constitutes all of the \ac{DM} and is virialized in the galaxy with a characteristic momentum $k\simeq 10^{-3}m_a$. The goal of this section is to calculate solutions to the EOM that asymptote towards the virialized axion \ac{DM} background at spatial infinity.

Before we proceed with solving the EOM, it is important to mention that they are not valid for cases where the mass reduction dominates over the vacuum mass, i.e. $\dm^2>\ma^2$. 
This region of invalidity is depicted as the red region in \Fig{parspace}. 
Here, the axion field can be sourced by compact objects, independently of constituting significant parts of the \ac{DM}, which leads to interesting phenomenology \cite{Hook:2017psm,Zhang:2021mks,Balkin:2022qer,Balkin:2023xtr,Gomez-Banon:2024oux,Kumamoto:2024wjd}. To treat this region properly, one should consider fluctuations around the $r$-dependent axion vacuum expectation value and linearize the \acp{EOM} around it. Given the stringent constraints that arise by sourcing the axion, we do not pursue this approach in this work.

Since we are interested in steady-state solutions, we can expand the axion field as a sum over plane-waves weighted by the appropriate velocity-dependent factor $f_\omega$, i.e., $a(\vect{r},t) \simeq \sum_\omega f_\omega a_\omega(\vect{r}) e^{i\omega t+\phi_\omega}+h.c.$. Inserting this decomposition into the axion EOM, we obtain equations for each frequency mode,
\begin{align}
\label{eq:EOM2}
    \begin{cases}
        \left(\omega^2-m_a^2+\nabla^2\right)a_\omega=0 \,,&  r>R \\
        \left(\omega^2-m_a^2+\nabla^2+\dm^2\right)a_\omega=0 \,,& r<R
    \end{cases}\,.
\end{align}
The solutions of interest are those that, far from the massive body, converge to the standard \ac{DM} solution
\begin{align}\label{eq:AsympAxionDMSol}
a(\vect{r},t)\Big|_{r\gg R}\longrightarrow  \,2a_0 \cos{\left(\omega_a t- \vec{k}\cdot\vec{r}\right)}\,,
\end{align}
where $a_0=\sqrt{\rho_{\rm DM}/(2m_a^2)}$ (this definition is half that of the more common notation for $a_0$). 
In what follows, we factor $e^{i\omega t}$ out of the solution, because the axion energy remains unchanged across the matter-vacuum boundary. Hence the field value is normalized to $a_0e^{-i\abs{\vec{k}}\abs{\vec{r}}\cos(\theta)}$  at $\abs{\vec{r}}\to\infty$, where $\theta$ is defined as the angle between $\vec{k}$ and $\vec{r}$. 
We omit the random phase at infinity for simplicity, see appendix~\ref{app:virial} for the more complete treatment of observables for a non-monochromatic field. 
 
General solutions to \Eq{eq:EOM2} can be found outside ($r>R$) and inside ($r<R$) the massive body, with the boundary conditions at $r=R$ fixing the unknown coefficients. 
The solution outside has to be composed of the incoming wave, i.e., the solution at $r\rightarrow\infty$, and outgoing scattered waves, which vanish at $r\rightarrow\infty$ in order to satisfy the \ac{DM} boundary condition. 
Given the spherical symmetry of the problem, angular momentum is conserved. Furthermore, given the azimuthal symmetry of the boundary conditions around the momentum of the incoming wave $\vec{k}$, we know that our solutions decompose to $l$-modes with $m=0$ (dictated by the choice of the \ac{DM} boundary condition in \eqref{eq:AsympAxionDMSol}).  
Following these considerations, the solution outside the spherical body may be expressed solely in terms of the $m=0$ spherical harmonic, the spherical Bessel functions $j_\ell$, and spherical Hankel functions of the first kind $h^{(1)}_\ell$ for outgoing waves
\begin{align}
\label{eq:oursolout}
    a^{\rm out}_\omega\left(r,\theta\right)= \sum^\infty_{\ell=0} a^{\rm out}_{\omega,\ell}\left(r,\theta\right)  =  a_0\sum_{\ell=0}^{\infty} (2\ell + 1) \, i^\ell \, j_\ell(kr) \, P_\ell(\cos \theta) + \sum_{\ell=0}^{\infty} c_\ell^{\text{out}} \, h_\ell^{(1)}(kr) \, P_{\ell}(\cos\theta) \, , 
\end{align}
where $k^2=m_a^2-\omega_a^2$ is the axion momentum outside the massive body, and $P_{\ell}$ the Legendre polynomials of order $\ell$. 
The first term in \Eq{eq:oursolout} corresponds to the free \ac{DM} solution at $r\to\infty$, while the second term $\propto c_\ell^\text{out}$ describes the outgoing scattered wave (adding terms $\propto h_{\ell}^{(2)}$ would correspond to further incoming spherical waves), with amplitudes that vanish at $kr\rightarrow \infty$. 
Matching the solution and its first derivative at $r=R_\oplus$ to the solution inside the massive body fixes the coefficients $c_\ell^{\rm out}$. 

The solution inside the sphere may also be expanded in terms of Legendre polynomials and spherical Bessel functions. 
However, the solution must not diverge at any point inside the sphere, so only the $j_\ell$ spherical Bessel functions can contribute. We therefore obtain
\begin{align}
    \label{eq:oursolin}
    a^{\rm in}_\omega\left(r,\theta\right) = \sum^\infty_{\ell=0} a^{\rm in}_{\omega,\ell}\left(r,\theta\right)=\sum_{\ell=0}^{\infty} c_\ell^{\text{in}} \, j_\ell(qr) \, P_\ell(\cos \theta)\,.
\end{align}
Energy conservation, along with the fact that the oscillation frequency is unchanged in our steady-state solution, enforces that the momentum-squared
inside Earth is $q^2 = k^2+\dm^2>k^2$.\,\footnote{Here, $\dm^2>0$ following \Eq{eq:axion_nucleon_int}. If the coupling in \Eq{eq:axion_nucleon_int} had the opposite sign, then $\dm^2 < 0$ and the solutions for the scalar field would exhibit pure screening~\cite{Hees:2018fpg,Banerjee:2022sqg,Stadnik:2024highlighting,Burrage:2024mxn}.
As we focus on light QCD axions in this work, we do not consider this case in detail.} 
Matching term by term at the surface of the sphere, one finds
\begin{align}
\label{eq:al}
a^{\rm out}_{\omega,\ell}\left(r,\theta\right)&=a_0\frac{ i^{\ell} (2 \ell+1)P_\ell\leri{\cos\theta} }{Q_\ell} \left[ y_\ell(k r)\text{Im}\leri{Q_\ell} +j_\ell(k r)\text{Re}\leri{Q_\ell} \right] \, , \\
a^{\rm in}_{\omega,\ell}\left(r,\theta\right) &=-a_0\frac{i^{\ell} (2 \ell+1)P_\ell\leri{\cos\theta} }{k R^2 Q_\ell} j_\ell(q r)\,,\\
\text{Re}\leri{Q_\ell} &= k j_\ell\leri{q R} y_{\ell+1}\leri{k R}-q j_{\ell+1}
(q R) y_\ell\leri{k R}\,,\\
\text{Im}\leri{Q_\ell}& = q j_{\ell+1}
(q R) j_\ell\leri{k R}-k j_\ell\leri{q R} j_{\ell+1}\leri{k R}\,.
\end{align}
Here, we used spherical Bessel functions of the second kind $y_\ell$ instead of the spherical Hankel function, and the two are related by $h_\ell^{(1)}(z)\equiv j_\ell(z)+iy_\ell(z)$.
One can also write $Q_{\ell}=i[\kd j_{\ell+1}(\kd R)h_\ell^{(1)}(kR)-kj_\ell(\kd R)h_{\ell+1}^{(1)}(kR)]$ instead. 
 
Note that the scattered waves, or the deviation from the free case (i.e., the incident plane-wave solution), is given by
\begin{align}\label{eq:scatt}
    a_{\omega,\ell}^{\rm scattered} = a_0i^{\ell}\leri{2\ell+1}P_\ell\leri{\cos\theta}\left[y_\ell\leri{kr} -i j_{\ell}\leri{kr}\right]\frac{{\rm Im}\leri{Q_\ell}}{Q_\ell}\,.
\end{align}

When $k=q$, we obtain $\textrm{Im}(Q_\ell)=0\,, \textrm{Re}\leri{Q_\ell} = -1/\leri{kR^2}$\,, and the waves inside and outside the sphere are simply the unperturbed \ac{DM} solution, as expected. 
On the other hand, we expect a strong effect from a mode $\ell$ when the scattered wave introduces an order one modification over the free one.

In this work, to simplify our calculations, we often remove the angular variable by averaging upon it. Then, the averaged field-amplitude-squared is given by
\begin{equation}\label{eq:amp_averaged}
    \langle |a_\omega(R)|^2 \rangle_{_\theta} = \left( \frac{a_0}{k^2 R^2} \right)^2 \sum_{\ell=0}^\infty (2\ell + 1) \left| \frac{j_\ell(q R)}{h_{\ell+1}^{(1)}(k R)\, j_\ell(q R) - \left( \frac{q}{k} \right) h_\ell^{(1)}(k R)\, j_{\ell+1}(q R)} \right|^2 \ ,
\end{equation}
where we introduce the notation $\langle \cdot \rangle_{_\theta}$ to indicate that the quantity has been averaged over $\theta$. The full field power spectral density is $\langle |a_\omega|^2\rangle_\theta+\langle |a_{-\omega}|^2\rangle_\theta =2\langle |a_\omega|^2\rangle_\theta$.

\begin{figure}[h!]
    \centering
    \includegraphics[width=0.45\linewidth]{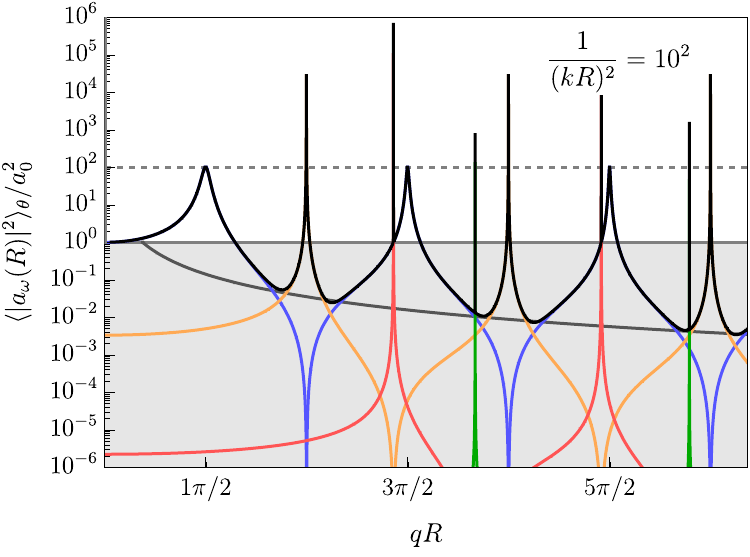}
    \quad\includegraphics[width=0.45\linewidth]{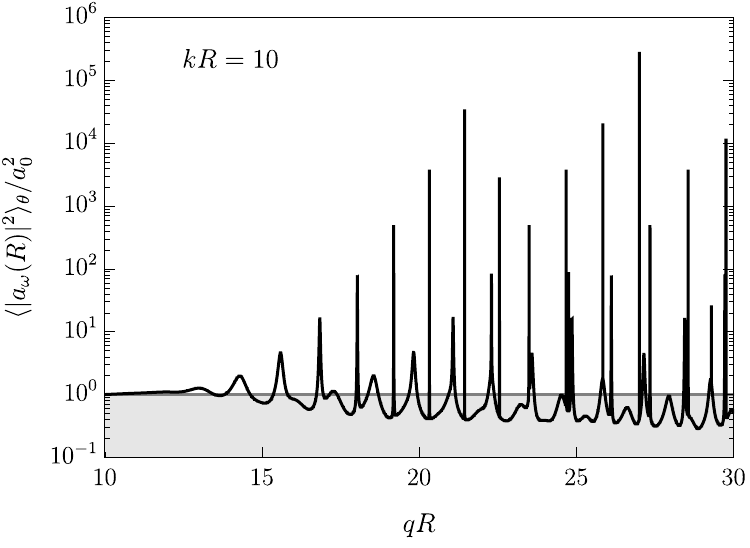}
    \includegraphics[width=0.45\linewidth]{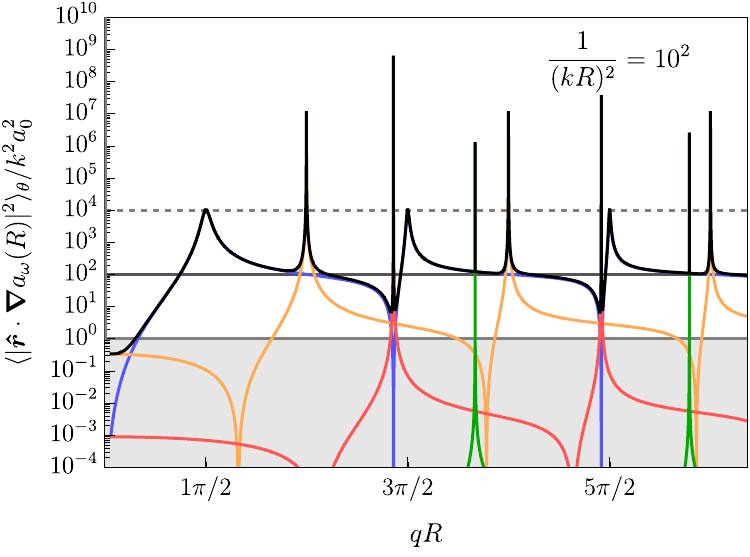}
    \quad\includegraphics[width=0.45\linewidth]{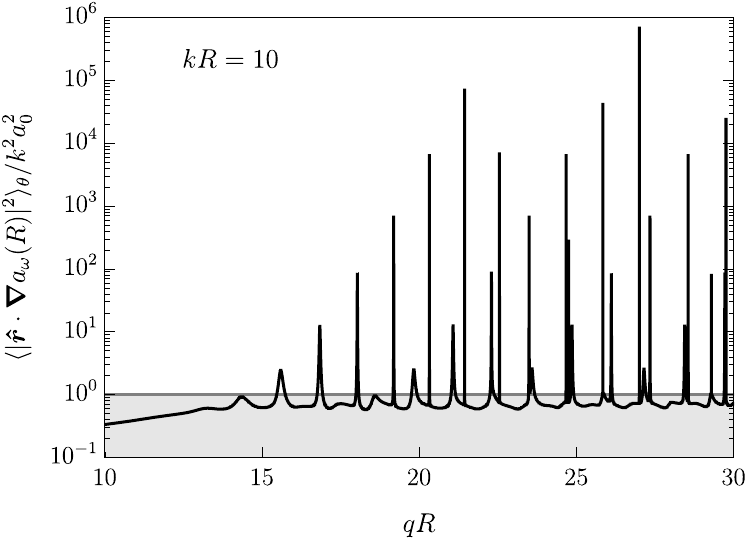}
    \includegraphics[width=0.45\linewidth]{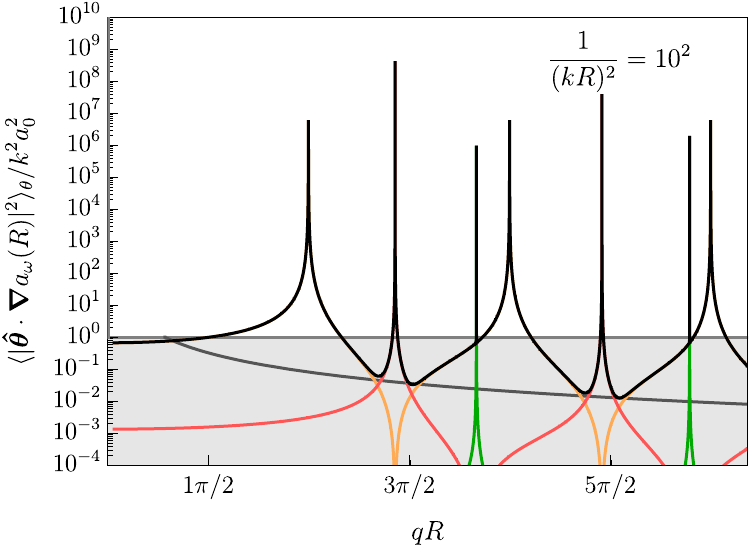}
    \quad\includegraphics[width=0.45\linewidth]{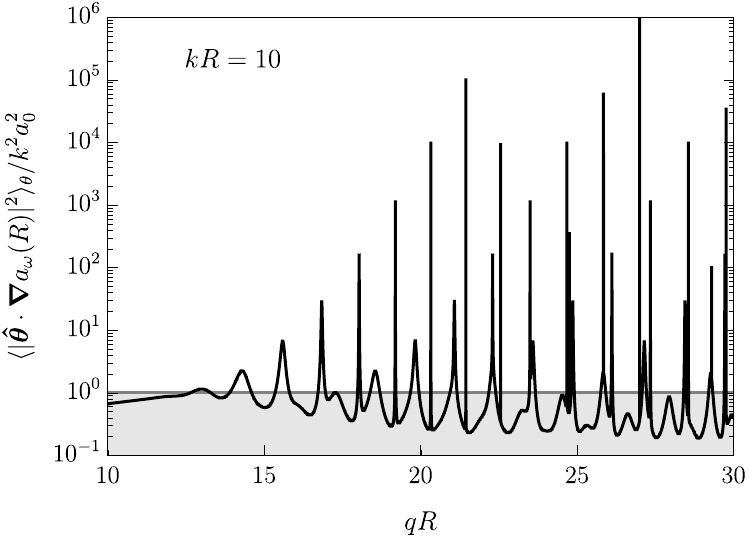}
    \caption{Upper (left and right) panels compare the angular average of the axion field value-squared with the na\"ive vacuum expectation of $\langle |a(R_\oplus)|^2\rangle = a_0^2$. 
    Middle panels compare the angular average of  $\vec{\hat{r}}\cdot\vec{\nabla} a(R_\oplus)$-squared with the na\"ive expectation of $\langle |\vec{\nabla} a(R_\oplus)|^2\rangle \sim k^2 a_0^2$.
    Bottom panels compare the angular average of $\vec{\hat{\theta}}\cdot\vec{\nabla} a(R_\oplus)$-squared with the na\"ive expectation of $\langle |\vec{\nabla} a(R_\oplus)|^2\rangle \sim k^2 a_0^2$. \textit{Left panels:} Summed solutions for $kR=0.1$ up to $\ell=3$. The full solution is shown in black, while single terms, i.e., $\ell=0,1,2,3$, are shown in blue, orange, red and green, respectively. See the text for explanations of the darker and dashed gray lines. 
    Note that the full solutions do not resolve the resonances up to their actual height. 
    While the the field amplitude and the $\theta$-component of the gradient (top and bottom panels) are suppressed by $\propto kR/(qR)^2$ off-resonance, the $r$-component of the gradient is enhanced by $1/(kR)^2$.
    \textit{Right panels:} The full solution for $kR=10$ summed up to $\ell=30$. 
    }
    \label{fig:l3sol}
\end{figure}

As can be seen from \Fig{l3sol}, in both regimes $kR>1$ or $kR<1$, resonances and suppressions occur for $q>k$, depending on the precise value of $qR$. 
For instance, in the $kR=0.1$ case (left panel of \Fig{l3sol}), the $\ell=0$ partial wave contribution exhibits a resonance at $qR= (2n-1)\pi/2$ with $n\in \mathbb{N}_+$, while zeroes are found at $qR\approx n\pi$, at which point the $\ell=1$ partial wave contribution exhibits a resonance. 
It should be noted that for monochromatic axions,  generically, the field value would be damped, as enhancements occur only in small regions.
In contrast to \cite{Bauer:2024hfv,Bauer:2024yow}, we find that solutions do not exhibit actual divergences once finite momentum boundary conditions are chosen. 
Still, for a monochromatic wave-packet with a finite momentum $k$, we find pronounced resonances of height $\sim (2\ell+1)^2y_{\ell}^2(kR)$, which scales as $\propto (kR)^{-2\ell-2}$ (with a proportionality factor that scales roughly as $\ell^{2\ell+2}$ for $\ell>kR$). 
Indeed, the relative height of the $\ell=0$ contribution is $10^2$ as indicated by the gray dashed line in \Fig{l3sol}, while that of the $\ell=1$ contribution is roughly $10^4$, and so on. 
The heights (and number) of higher $\ell$ resonances are not correctly displayed due to limited resolution. When integrating over virially distributed \ac{DM}, the extremely narrow width of the resonances regulates their height (more precisely, the width scales with powers of $k R$ and $\ell$ that are inverse to those that determine the height, see Appendix~\ref{app:resonances}). 
Furthemore, away from resonance in the $kR<1$ case we find a suppression of $2\sqrt{3}kR/(qR)^2$, which is shown by the darker gray line in the top left panel of \Fig{l3sol} for $kR=0.1$. 

Similarly, we focus on comparing the $r$- and $\theta$-components of the $\theta$-averaged axion gradient squared with the naive expectation, $\langle|\vec{\nabla}a|^2\rangle \sim k^2 a_0^2$. 
The analytical results are readily found as
\begin{subequations}
\begin{align}
    \frac{\langle |\vec{\hat{r}}\cdot\vec{\nabla}a_\omega(R)|^2 \rangle_{_\theta}}{k^2 a_0^2}=&\left(\frac{1}{kR}\right)^6\sum_{\ell=0}^\infty(2\ell+1)\left|\frac{\ell j_\ell(qR)-qRj_{\ell+1}(qR)}{h_{\ell+1}^{(1)}(k R)\, j_\ell(q R) - \left( \frac{q}{k} \right) h_\ell^{(1)}(k R)\, j_{\ell+1}(q R)}\right|^2,\\
    \frac{\langle |\vec{\hat{\theta}}\cdot\vec{\nabla}a_\omega(R)|^2 \rangle_{_\theta}}{k^2 a_0^2}=&\left(\frac{1}{kR}\right)^6\sum_{\ell=0}^\infty(2\ell+1)\ell(\ell+1)\left|\frac{j_\ell(q R)}{h_{\ell+1}^{(1)}(k R)\, j_\ell(q R) - \left( \frac{q}{k} \right) h_\ell^{(1)}(k R)\, j_{\ell+1}(q R)}\right|^2,
\end{align}
\end{subequations}
and shown in the middle and lower panels of \Fig{l3sol} for $kR=0.1$ as well as $kR=10$. In the case of $kR<1$, we find that the relative enhancement of the radial component of the axion gradient squared is enhanced by $1/(kR)^{4}$ for the $\ell=0$ mode.
In contrast to the amplitude squared, for large $qR$ and off-resonance, we find a relative asymptotic enhancement of $1/(kR)^2$, which is indicated by the darker gray line in the middle-left panel of \Fig{l3sol}.
For the $\theta$ component of the gradient it is worth mentioning that the $\ell=0$ mode does not contribute and that, similarly to the amplitude square, one finds a suppression $\propto kR/(qR)^2$ away from resonance, as indicated by the darker gray line. 
Finally, we observe that in the limit $q \to k$, neither the $r$-component nor the $\theta$-component asymptotes to unity. This behavior is expected, as only the full squared gradient converges in this limit, specifically $\langle |\vec{\nabla}a_\omega(R)|^2 \rangle_{_\theta} \to k^2 a_0^2$ for $q \to k$.

In this section, we focused on solutions with a well-defined ${\bf k}$ at infinity. However, within the Standard Halo Model, the \ac{DM} velocity is a stochastic variable. The appropriate treatment of this is discussed in Appendix~\ref{app:virial}. However, if the field profile is not strongly $k$ dependent, this integration is not important. For $\kvir R\ll 1$, only rarely do resonances play a role, and when they do not, the monopole $(\ell=0)$ and dipole $(\ell=1)$ contributions dominate. Hence, in the following section, we focus on these contributions, which are expected to be most important for $\kvir R\ll 1$, and already present a rich and interesting phenomenology.

\section{Monopole and Dipole Contributions}
\label{sec:lowL}

In this section, we focus on the monopole $\ell=0$ and dipole $\ell=1$ contributions. 
In the case when $kR\gg 1$, we find that many $\ell$-modes contribute, and one must sum up to $\ell=kR+\mathcal{O}(1)(kR)^{1/3}$ (as can be understood by using the asymptotic spherical Bessel functions formulae~\cite{besselcite}) to estimate the non-resonant contributions. 
For $kR\ll 1$, we can Taylor expand our solution and its derivative around $k\to 0$ (away from resonances) to find,
\begin{equation}
\begin{aligned}
&\lim_{k\to 0}a_{\omega,\ell}(R)=
a_0\frac{(1+2\ell)\(-\frac{ikR}{2}\)^\ell\Gamma\(\frac{1}{2}-\ell\)}{\sqrt\pi}\frac{j_\ell(\dm R)}{\dm R \, j_{\ell-1}(\dm R)}P_\ell(\cos\theta) \ , \\
&\lim_{k\to 0}\partial_ra_{\omega,\ell}(R)=-a_0\frac{(1+2\ell)\(-\frac{ikR}{2}\)^\ell\Gamma\(\frac{1}{2}-\ell\)}{\sqrt\pi}\frac{\[(1+\ell)j_\ell(\dm R)-\dm Rj_{\ell-1}(\dm R)\]}{\dm \, R  \, j_{\ell-1}(\dm R)}P_\ell(\cos\theta) \ .
\end{aligned}
\end{equation}

Due to the $(kR)^\ell$ scaling, one would expect that for $kR<1$ one may focus on the low $\ell$ modes. Hence, this section which focuses on $\ell=0,1$ can be understood as describing the axion in the $\kvir R\ll 1$ regime. Interestingly, the $\ell=0,1$ modes generically dominate the overall amplitude in \Eq{eq:amp_averaged}, even when sampling $k$ stochastically (see Appendix~\ref{app:virial}). 
While one may worry about the presence of resonances from high $\ell$ modes, for $\kvir R\ll 1$, this rarely occurs. 
One may estimate the distance in $q$ between two adjacent resonances for large $qR$ to be roughly $1/(qR^2)$. 
Since the virial integral integrates over a region $ \dm\leq  q\lesssim \sqrt{\dm ^2 +{\kvir^2}}$, one can see that the region being scanned is narrower than $1/(qR^2)$ for $\kvir R<1<qR$. 
Hence, one should not expect resonances to contribute. 
For further discussion, see also Appendix~\ref{app:resonances}. 

At the end of the section, we will also discuss more broadly the field value and gradient deep inside Earth, and far outside of it.

\subsection{Terrestrial $a$ observables}

Let us first discuss the field values of the $\ell=0,1$ modes at $r=R$.
We begin with $\ell=0$. The field value for $\ell=0$ at the boundary of the object is 
\begin{equation}
a_{\omega,0}(R)=\frac{e^{-ikR}a_0}{-ikR+{\kd}R\cot(\kd R)}\,.
\label{eq:a0}
\end{equation}
This is to be compared to the $\ell=0$ component of the unperturbed axion wave, which is simply $a_0\sin(k R)/(k R)$, which is approximately $a_0$ for $kR\ll 1$. As $\kd R$ grows, the modified field value would drastically change, since $\cot(\kd R)$ diverges when $\sin(\kd R)=0$. 
In fact, unless $\cos(\kd R)=0$, for increasing values of $\kd R$, the $\ell=0$ mode would be suppressed by $1/(\kd R)$ compared to the expectation far from Earth. 
This means that unless $\cos(\kd R)\approx 0$, where the field is enhanced by $1/(kR)$ (which corresponds to the $qR=\pi n+\pi/2$ resonances displayed in Fig.~\ref{fig:l3sol}), the $\ell=0$ mode would be suppressed for $\kd R\gg 1$, i.e., for $f_a \ll (f_a)_\oplus^c$ [where the critical decay constant $(f_a)^c_\oplus$ is given by Eq.~\eqref{eq:EarthCritFa} for Earth]. 
Away from resonances, the axion field value in the large $q R$ limit becomes approximately
\begin{align}
    a_{\omega, 0}(R) \sim a_0 \min \left[1, \frac{f_a}{(f_a)_\oplus^c} \right] \ .
    \label{eq:axionMatter}
\end{align}

Now let us examine how the $\ell=1$ mode changes the picture. 
The field value at the surface of Earth is

\begin{equation}
\label{eq:a1}
a_{\omega,1}(R,\theta)=
\frac{
    3a_0 k R e^{-i k R} \cos(\theta)\left[q R \cos(q R) - \sin(q R)\right]
}{
    \left[kR (qR)^2 - i R^2 (k^2 - q^2) - 3 k R + 3\right] \sin(q R)
    - q R \left[ 3-kR(3-ikR)\right] \cos(q R)
}
\, .
\end{equation}
For  $kR\lesssim 1\lesssim \kd R$ and away from a resonance or anti-resonance [i.e., when $j_{0,1}(qR)=0$ respectively], this field is suppressed by the factor $\mathcal{O}(k/\kd)$ compared to the amplitude of the $\ell=0$ mode. 
However, when the $\ell=0$ mode is suppressed (i.e., when $\sin(\kd R)=0$), this mode becomes enhanced as $1/(kR)$ (which is again evident in Fig.~\ref{fig:l3sol}). 
This means that within the aforementioned region, for most choices of $\kd R$, this mode is negligible; however, for specific choices of $\kd R$, where the $\ell=0$ mode is suppressed, this mode becomes enhanced.

\subsection{Terrestrial $\partial a$ observables}
 
Ignoring the effect of the massive body, the gradient is of order $ka_0$. 
However, the existence of a matter-sourced potential can lead to a significant deviation from this expectation.
Since $P_0(x)=1$, the $\ell=0$ mode has only a radial derivative, whose value at the surface of the body reads
\begin{equation}
\partial_ra_{\omega,0}(R)=\frac{e^{-ikR}a_0}{R}\frac{\kd R\cos(\kd R)-\sin(\kd R)}{\kd R\cos(\kd R)-ikR\sin(\kd R)}\,.
\label{eq:da0}
\end{equation}
For $\kd R\gtrsim 1$, its magnitude is generically enhanced with respect to the $ka_0$ expectation (i.e., with respect to the $\dm=0$ case) by a factor of $1/(kR)$, and even by $1/(kR)^2$ on the $\ell=0$ resonance, i.e., when $\cos(\kd R)=0$. We emphasize that this enhancement is compared to $ka_0=(ka)_{\rm no~Earth}$, so that even though the field values $a(R)$ become smaller for increasing $\kd R$, the derivatives are enhanced in absolute terms. 

We find that the typical value of the radial gradient away from any resonances, given in terms of the axion mass and coupling constant is
\begin{align}
    \partial_r a_{\omega}(R_\oplus) \sim k a_0 \left(1+\max  \left[1,\frac{1}{k R_\oplus}\right]\,\min \left[1, \left(\frac{(f_a)^c_\oplus}{f_a}\right)^2 \right]\right) \ ,
    \label{eq:radialGradientEarth}
\end{align}
with $(f_a)^c_\oplus$ defined for Earth in Eq.~\eqref{eq:EarthCritFa}.
We have written this with an overall factor of $k a_0$, which is the naive solution when not accounting for matter effects, so as to highlight the enhancement in the small $kR_\oplus$ limit. 

The $\ell=1$ term has a radial derivative and one in $\hat{\theta}$. Focusing again on $k R\lesssim 1\lesssim \kd R$, when $\sin(\kd R)=0$, both derivatives scale like $1/(kR^2)$, while for other values of $\kd R$, the radial derivative scales like $k$, while the gradient in the angular direction scales like $k/(\kd R)$. 

Interestingly, even for $qR\lesssim1$, $\nabla a$ can still have a significant enhancement compared to the vacuum case. For $kR<qR<1$, $\boldsymbol{\nabla} a\approx -a_0\dm^2R{\boldsymbol{\hat{r}}}/3$, which can be far larger than $ka_0$, which is the vacuum prediction. This is different from the sourced soliton solution, which stops being important for $\dm R<1$.

\subsection{Field profiles away from Earth's surface}

\begin{figure}[h!]
    \centering
    \includegraphics[width=0.45\linewidth]{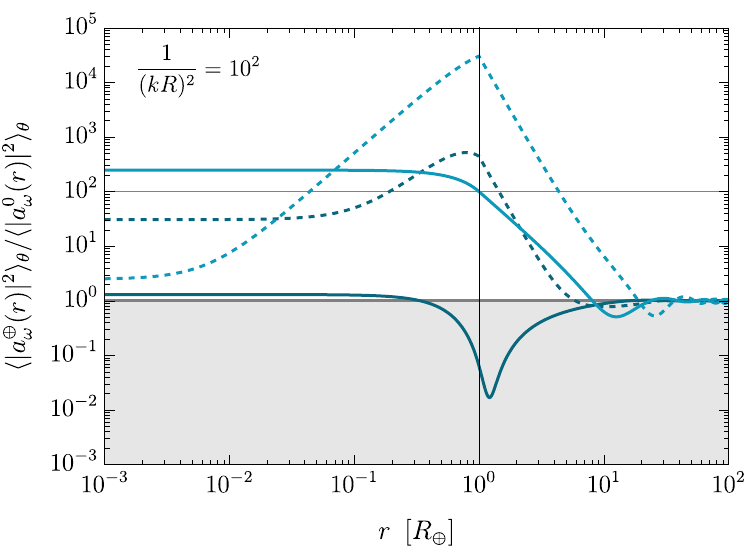}
    \quad\includegraphics[width=0.45\linewidth]{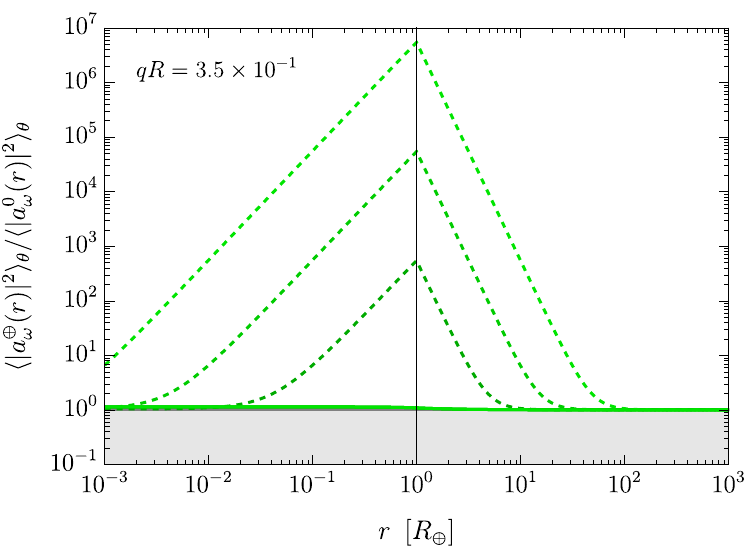}
    \includegraphics[width=0.45\linewidth]{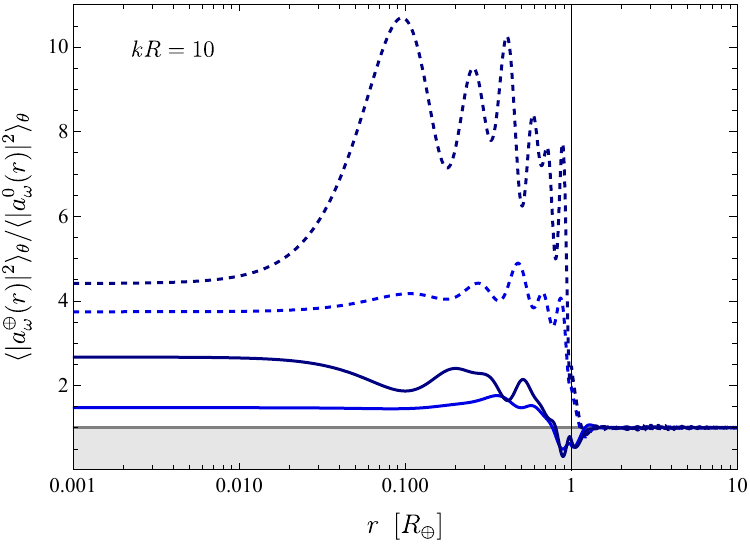}
    \includegraphics[width=0.45\linewidth]{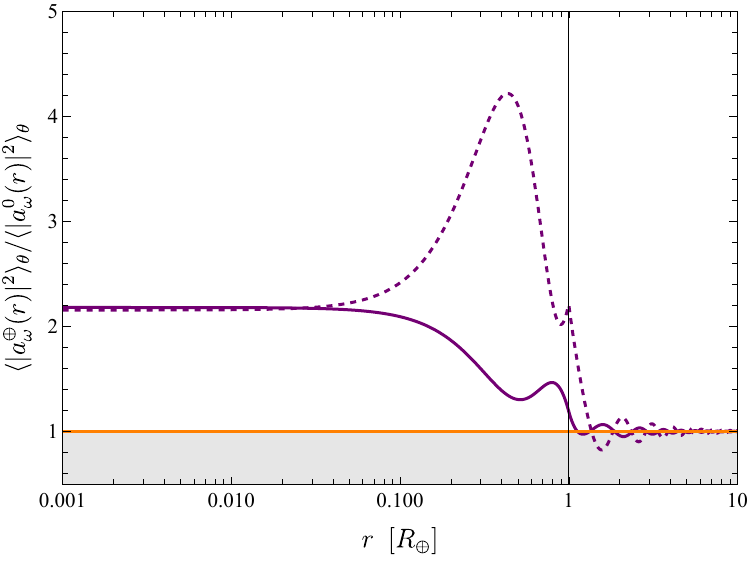}
    \caption{\textit{Top panels:} Solutions for the field amplitude (colored solid lines) and the gradient  (colored dashed lines) as a function of $r$ for $kR<1$. Top left lies in the green-blue region of \Fig{parspace} and corresponds to $kR=0.1$ with an on-resonance solution ($qR=\pi/2$ -- see \Fig{l3sol}) shown in lighter blue-green, and an off-resonance contribution ($qR=\pi/2+1/2$ -- see \Fig{l3sol}) shown in darker blue-green.
    The top right panel shows solutions in the pure green region of \Fig{parspace} with fixed $qR\simeq3.5\times10^{-1}$ ($f_a=10^{14}~\mathrm{GeV}$). 
    From dark green to light green we vary the axion mass $m_a=10^{-13}~\mathrm{eV},10^{-14}~\mathrm{eV},10^{-15}~\mathrm{eV}$.
    \textit{Bottom panels:} Solutions for the field amplitude (colored solid lines) and the gradient  (dashed lines) as a function of $r$ for $kR>1$. $\mathcal{O}(1)$ enhancements can be found within Earth, see text for details.
    Notably, in the bottom-right panel, we plot an orange line corresponding to the \emph{canonical} QCD axion, which is unaffected by Earth's presence, as argued in the text. 
    See text for the specification of all parameters chosen in the plot.
    }
    \label{fig:profiles}
\end{figure}

In this subsection, we investigate the axion amplitude and its gradient away from Earth's surface. 
This could significantly impact existing or future experiments away from the surface of Earth.
In the $kR<1$ limit, the solutions are again completely dominated by the $\ell=0$ and $\ell=1$ contributions.
In \Fig{profiles}, we compare our solutions for the field amplitude and the gradient with the asymptotic expectation away from Earth.
In particular, we compare the $\theta$-averaged amplitude and the $r$-component of the gradient to the free solution (and gradient thereof), see \Eq{eq:AsympAxionDMSol}.
The $\theta$-component of the gradient is directly proportional to the amplitude but without the $\ell=0$ contribution.

In the top-left panel, we show solutions for the field amplitude and the gradient for $kR=0.1$ (equivalent to $m_a\simeq3.1\times10^{-12}\,\mathrm{eV}$) for two values of $qR$, both in the intersection of the green and blue regions of \Fig{parspace}. 
In lighter turquoise we show the first $\ell=0$ resonance which occurs at $qR=\pi/2$ (see \Fig{l3sol}), corresponding to $f_a\simeq 2.2\times 10^{13}\,\mathrm{GeV}$.
As can be seen, at $r=R_\oplus$, we confirm the expected $1/(kR)^2=10^2$ enhancement as indicated by the vertical black line, while the squared gradient is roughly enhanced by $1/(kR)^4$ at the surface.
For $r<R_\oplus$, the field amplitude squared remains enhanced by $\sim 1/(kR)^2$ until the very center of Earth. 
For $r>R_\oplus$, the enhancement drops off within a distance of $\sim (1/k)=10R_\oplus$ and approaches the asymptotic \ac{DM} solution, see \Eq{eq:AsympAxionDMSol}.
The gradient is maximally enhanced at $r=R_\oplus$ and drops off away from the surface. 
It asymptotes to the free solution outside of Earth, while a residual enhancement remains towards the center of Earth but fully ceases to exist as one approaches $r=0$. 
In darker turquoise, we show the amplitude and gradient off resonance, namely at $qR=\pi/2+1/2$, which corresponds to $f_a=9.6\times 10^{12}\,\mathrm{GeV}$.
As expected, the field amplitude is strongly suppressed at the surface of Earth, while the gradient is enhanced. 
Away from the surface, the field amplitude asymptotes towards the free solution.
While the gradient outside of Earth approaches the free solution, it remains enhanced inside for the radii shown, which arises due to the contribution from the $\ell=1$ mode for the range shown. 
At the origin, the gradient is no longer enhanced for the parameters shown. More generally, at the origin, the $\ell=0$ mode is the only one contributing to the field value, while the $\ell=1$ mode is the only one contributing to the gradient. Highly non-trivial enhancements can be had at the origin for different parameter choices, though as it is not clear that such enhancements have any measurable effect, we do not focus on classifying them.

In the top-right panel, we show solutions from the pure green region in~\Fig{parspace}, in which the gradient gets strongly enhanced, while the amplitude does not significantly change.
The solutions shown correspond to $f_a=10^{14}\,\mathrm{GeV}$ and varying axion masses $m_a=10^{-13}\,\mathrm{eV},\,10^{-14}\,\mathrm{eV},\,10^{-15}\,\mathrm{eV}$, which correspond to increasingly lighter shades of green, respectively. 
These correspond to the sets of values $kR_\oplus\simeq 3.2\times 10^{-3},3.2\times 10^{-4},3.2\times 10^{-5}$, while $qR$ being dominated by $\dm^2$ is hence $qR_\oplus\simeq 3.5\times10^{-1}$ for all values of $m_a$. 
Gradients at the surface are dramatically enhanced by a factor $\sim(qR)^2/kR$. 
This factor can be found from the simultaneous small $kR,~qR$ limit of Eq.~\eqref{eq:da0}, and is observed in the middle-left panel of  \Fig{l3sol}.
Away from Earth, the enhancement disappears away from the surface, depending on $m_a$. 

In the bottom-left panel, we show solutions from the pure blue region of Fig.~\ref{fig:parspace}. 
We show the amplitude and gradient for $kR=10$ and both $qR=16.5$ and $qR=20$ (blue and darker blue respectively), which corresponds to $m_a\simeq3.1\times10^{-10}\,\mathrm{eV}$ and $f_a\simeq2.6\times10^{12}\,\mathrm{GeV},\,2.0\times10^{12}\,\mathrm{GeV}$, respectively. 
Here it is no longer true that the $\ell=0,1$ contributions dominate and in the solutions shown, we have summed up to $\ell=100$. 
For both sets of parameters, we find an $\mathcal{O}(1)$ enhancement in both the gradient and amplitude values inside Earth. 
The gradient is also enhanced at the surface of Earth (vertical black line), but the field amplitude is suppressed slightly. 
Outside of Earth, the effects disappear at the scale proportional to the de-Broglie length of the axion i.e. $\propto (kR_\oplus)^{-1} R_\oplus$. 

In the bottom-right panel, in purple solid and purple dashed, we show a solution for $kR>1$ just outside the blue region, namely for $(m_a,f_a)=(10^{-10}~\mathrm{eV},10^{13}~\mathrm{GeV})$, which corresponds to $(kR,qR)\simeq (3.2,4.7)$.
While the effect at the surface of Earth is rather minor, $\mathcal{O}(1)$ effects can arise within Earth, both for the field amplitude and the gradient.
We would like to stress that \Fig{parspace} focuses on effects at the surface of Earth, which implies that effects within Earth can be found just outside of the regions shown.
In orange, we show solutions for the \emph{canonical} QCD axion with $kR=1$ for $f_a=10^{17}~\mathrm{GeV}$. 
As expected, no effect is found.

\section{Experimental Implications}
\label{sec:ExpConseq}

Our findings have important implications for experiments searching for axion \ac{DM} if they are only sensitive to the region of \Fig{parspace} where matter effects dominate the solution to the axion \ac{EOM}. Additional effects not discussed are new forces~\cite{Tilburg:2024wake,Day:2024blowing,Zhou:2025wax}, which may be observable, for example in equivalence principle tests~\cite{Berge:2017ovy,MICROSCOPE:2022doy,Hees:2018fpg,Banerjee:2022sqg}.

\subsection{Experiments targeting the axion wind}

\begin{figure}
    \centering
    \includegraphics[width = 0.8\linewidth]{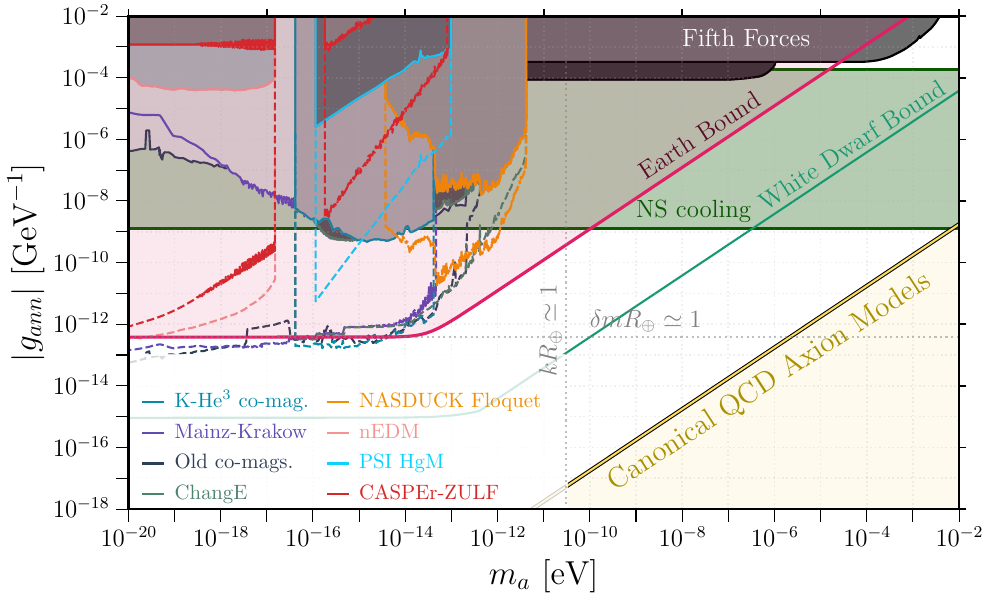}
    \caption{Possible enhancement of a selection of experiments sensitive to the axion gradient coupling to neutrons, $g_{ann}$, with $C_n = 10$. The grey shaded regions show the existing experimental constraints from the K-$^3$He co-magnetometer~\cite{Lee:2022vvb} (blue outline), the Mainz-Krakow co-magnetometers~\cite{Gavilan-Martin:2024nlo} (purple),  estimated sensitivity of old co-magnetometer data~\cite{Bloch:2019lcy} (charcoal), ChangE~\cite{Wei:2023rzs} (green),  NASDUCK~\cite{Bloch:2021vnn,Bloch:2022kjm} (orange), neutron EDM searches~\cite{Abel:2017rtm} (pink), $^{199}\text{Hg}$ precession~\cite{Abel:2022vfg} (cyan), and CASPEr-ZULF~\cite{Garcon:2019inh,Wu:2019exd} (red) under the vacuum axion assumption. 
    We show dashed lines with the same colour scheme the possible new constraints if their data were taken in such a way as to be sensitive to the radial gradient of the axion dark matter field. 
    Any possible new constraint within the red shaded region labeled ``Earth Bound'' is superseded by the Earth Bound \cite{Hook:2017psm} itself, as the dynamics leading to the enhanced sensitivity to $g_{ann}$ also leads to this exclusion limit. 
    The gray line shows the bound from White Dwarfs \cite{Balkin:2022qer}.
    In green, we show the bound from neutron star cooling \cite{Buschmann:2021juv,Springmann:2024mjp}.}
    \label{fig:GradExps}
\end{figure}

The axion-fermion coupling is defined according to the Lagrangian term $\mathcal{L}\sim g_{a\psi\psi}(\partial_\mu a)\bar \psi \gamma^5\gamma^{\mu}\psi$, with the coupling strength related to the decay constant $f_a$ through,
\begin{align}
\label{Eq:coupling_constant_definition}
g_{a\psi\psi}=\frac{C_\psi}{f_a},
\end{align}
with $C_\psi$ being a model-dependent and fermion-dependent parameter. 

Assuming non-relativistic fermions and a non-relativistic background field of axions (i.e., with a large occupation number), the fermion-axion interaction induces a splitting between spins aligned and anti-aligned with the velocity of the axions~\cite{OldWind},
\begin{align}
H_{a\psi\psi}=2g_{a\psi\psi}({\vec{\nabla}} a)\cdot {\bf S}_\psi=2g_{a\psi\psi}\sqrt{2\rho_{\rm DM}}{\rm sin}(\omega t-{\bf k}\cdot  {\bf r}){\bf v}\cdot {\bf S}_{\psi} \, . 
\end{align}
The above interaction is entirely analogous to the Zeeman splitting, induced in the presence of magnetic fields $H=\gamma_{\psi} {\bf B}\cdot {\bf S}_{\psi}$, with  ${\bf B}$ as the magnetic field, and $\gamma_{\psi}$ as the gyromagnetic ratio. Hence, magnetometers which are sensitive to the Zeeman splitting should also be sensitive to the presence of axions. Over the last decade, multiple experiments which utilize such methods have been proposed or performed~\cite{Bloch:2019lcy,Bloch:2021vnn,Bloch:2022kjm,Gavilan-Martin:2024nlo,Wu:2019exd,Garcon:2019inh,Wei:2023rzs,Xu:2023vfn,Jiang:2021dby,Flambaum_Patras_2013,Stadnik:2013raa,Stadnik_2017_book-thesis,Abel:2017rtm,Abel:2022vfg,Lee:2022vvb,Zhang:2023qmu,Bloch:2024uqb}. 
Such experiments tend to focus on the lower end of the frequency range, and hence, should the quadratic interaction of axions exist, this work could have large implications on such experiments, enhancing their reach, should they be sensitive to radial gradients. 

Currently, most experiments do not account for the enhancement calculated here, and hence lie deep within the region where the potential for the axion is inverted inside Earth (assuming $C_\psi\lesssim 1$). 
However, for sufficiently small axion masses, the enhancement of $\hat{r}\cdot \nabla a(R_\oplus)$ could be large enough that existing experiments probe some parameter space beyond the Earth bound. 
This is shown in Fig.~\ref{fig:GradExps}, where we have focused on the axion coupling to neutrons for illustrative purposes, and fixed $C_n = 10$. All existing experimental constraints assume the vacuum axion field value, and are consequently well inside the region of parameter space excluded by the Earth sourcing bound. However, if these experiments are actually sensitive to the radial gradient of the axion field, taking matter effects into account dramatically improves their sensitivity. We see in particular that for the smallest axion masses, the improvement can be as large as nine orders of magnitude. The improvement is less dramatic as one approaches the $kR_\oplus \simeq  1$ regime (vertical dashed grey line), at which point the enhancement is expected to disappear. We further see that enhancements are decreased as the sensitivity reaches the regime corresponding to $\dm R_\oplus \lesssim 1$ (horizontal dashed gray line). This is expected based on the form of Eq.~\eqref{eq:radialGradientEarth}, as in this regime, the factor $((f_a)^c_\oplus/f_a)^2 \lesssim 1$ decreases the enhancement from its peak value of $1/kR_\oplus$.

Note that our optimistic recast of the various axion gradient experimental sensitivities leads to new bounds that barely extend beyond the ``Earth Bound'' region (red in Fig.~\ref{fig:GradExps}). In that region, our calculation of the axion gradient about its \ac{DM} vacuum value is likely to be subdominant compared to the effect from the axion vacuum being shifted by Earth from $a= 0$ to $a=\pi f_a$~\cite{Hook:2017psm}. 
The coupling that leads to the enhanced $\grad a$ in our analysis is the same that leads to the shift in the axion vacuum. Therefore, benefiting from the enhanced gradient in the red-shaded region of Fig.~\ref{fig:GradExps} is likely to be impossible without being simultaneously excluded by the Earth Bound. The existence of White Dwarfs excludes the region above the light green line~\cite{Balkin:2022qer}. The coupling excluding this parameter space is also the same as that which leads to an enhanced $\grad a$ on Earth. Therefore, without significant model-building to avoid this bound, it is likely that enhancements of $\grad a$ do not lead to new constraints on QCD axions from existing experiments.

\subsection{Experiments targeting the axion-photon coupling}

The axion-photon coupling is defined according to the Lagrangian term $\mathcal{L} \sim g_{a\gamma\gamma}\,a\,F\,\tilde{F}$, with the coupling strength related to the decay constant $f_a$ through
\begin{align}
    g_{a\gamma\gamma} = C_\gamma \frac{\alpha}{2\pi f_a} \ ,
\end{align}
where $C_\gamma$ is a model-dependent numerical prefactor that is typically $\mathcal{O}(1)$. 
As this coupling involves a vertex of two photons and one axion, a \ac{DM} search targeting this interaction usually involves the application of a background  \ac{EM} field. Typically, a current-carrying wire of length $\lw$ produces that field within a cavity of length-scale $\ell$  (we here focus on experiments utilizing electromagnets rather than permanent magnets). The signal for such experiments is the conversion of the axion to an \ac{EM} field with a frequency shifted by the axion frequency $\omega$ compared to the background. We can divide such experiments according to whether the applied background  \ac{EM} field is DC or AC. Existing and proposed experiments searching for axion conversion in a DC background (magnetic) field include traditional haloscopes~\cite{Sikivie_1983,ADMX:2024xbv,CAPP:2024dtx} and lumped LC resonators~\cite{Kahn:2016aff,Ouellet:2018beu,Salemi:2021gck,DMRadio:2022pkf,DMRadio:2022jfv,DMRadio:2023igr}. Existing and proposed experiments searching for axion conversion in an AC background  \ac{EM} field include Heterodyne SRF~\cite{Berlin:2019ahk,Lasenby:2019prg,Berlin:2020vrk,Giaccone:2022pke} and ADBC~\cite{Liu:2018icu,Pandey:2024dcd}.

To estimate the effects of our findings on experimental searches, we use the wave equations for the EM fields to first order in the axion coupling $g_{a\gamma\gamma}$. A more complete discussion of their origin is found in Appendix~\ref{app:Vpt2}. The $E$- and $B$-field equations of motion are
\begin{align}
    \label{eq:E1eom}
    (\nabla^2 -\partial_t^2) \vect{E}_1 &= -g_{a\gamma\gamma} \left( \grad\left(\vect{B}_0 \cdot \grad a \right) +\partial_t\left(\vect{E}_0 \times \grad a - \partial_t a \,\vect{B}_0 \right)\right) \ , \\
    (\nabla^2 -\partial_t^2) \vect{B}_1 &= g_{a\gamma\gamma}\nabla \times \left(\vect{E}_0 \times \grad a - \partial_t a \,\vect{B}_0 \right)  \ ,
    \label{eq:B1eom}
\end{align}
which we will now give approximate solutions to by na\"ive dimensional analysis.

\subsubsection{Experiments in the Magneto-quasistatic limit}

Our findings indicate that axion amplitudes and gradients are only affected for small axions or large couplings (see Fig.~\ref{fig:parspace}). Therefore, the experiments with DC background  \ac{EM} fields targeting this parameter space typically operate in a regime where there exists a  hierarchy between the typical length scale of the apparatus $l$ and the frequency of the signal $\omega$, such that $\omega l \ll 1$. We further assume they are operated with a background DC magnetic field $B_0$ and no background electric field.

As long as the axion gradient in the radial direction dominates, the equations for the axion-induced  \ac{EM} fields can be roughly approximated as
\begin{align}
    \label{eq:E1mqs}
    \frac{1}{l^2}E_1 &\sim g_{a\gamma\gamma} B_0\left(m_a^2 a - \frac{1}{l} \partial_r a - \partial_r^2 a \right) \ , \\
    \frac{1}{l^2}B_1 &\sim - g_{a\gamma\gamma}B_0 m_a\left(\frac{1}{l} a + 
\partial_r a\right)\ .
    \label{eq:B1mqs}
\end{align}
Neglecting matter effects, $\partial_r a \sim k a_0$, and we get the usual solution for $E_1\sim (m_a l)^2 B_0 a_0$, while $B_1 \sim (m_a l) B_0 a_0$~\cite{Bogorad:2019pbu}. However, in the matter-affected region, either $\nabla a$ is affected alone (green coloured region of Fig.~\ref{fig:parspace}), or $a(R_\oplus)$ is also affected (blue coloured region of Fig.~\ref{fig:parspace}). Therefore, the common assumption that the axion-induced electric field is parametrically smaller than the induced magnetic field can be false. 
Using the results of Eqs.~~\eqref{eq:axionMatter} and~\eqref{eq:radialGradientEarth} for $a(R_\oplus)$ and $\partial_r a(R_\oplus)$ respectively, we find
\begin{align}
    \label{eq:E1mqsEarth}
    E_1(R_\oplus) &\sim g_{a\gamma\gamma} B_0  a_0 \left((l m_a)^2 \min \left[1, \frac{f_a}{(f_a)^c_\oplus} \right] - \left(\frac{l}{R_\oplus} - \frac{l^2}{R_\oplus^2}\right) \min \left[1, \left(\frac{(f_a)^c_\oplus}{f_a}\right)^2 \right] \right) \ , \\
    B_1(R_\oplus) &\sim -g_{a\gamma\gamma} B_0  m_a a_0 \left( l \min \left[1, \frac{f_a}{(f_a)^c_\oplus}  \right] + \frac{l^2}{R_\oplus}\min \left[1, \left(\frac{(f_a)^c_\oplus}{f_a}\right)^2 \right] \right)\ .
    \label{eq:B1mqsEarth}
\end{align}
We emphasize that these approximate relations are robust for $kR_\oplus \lesssim 1$, but might break down for $kR_\oplus \gtrsim 1$.

\begin{figure}
    \centering
    \includegraphics[width = 0.8\linewidth]{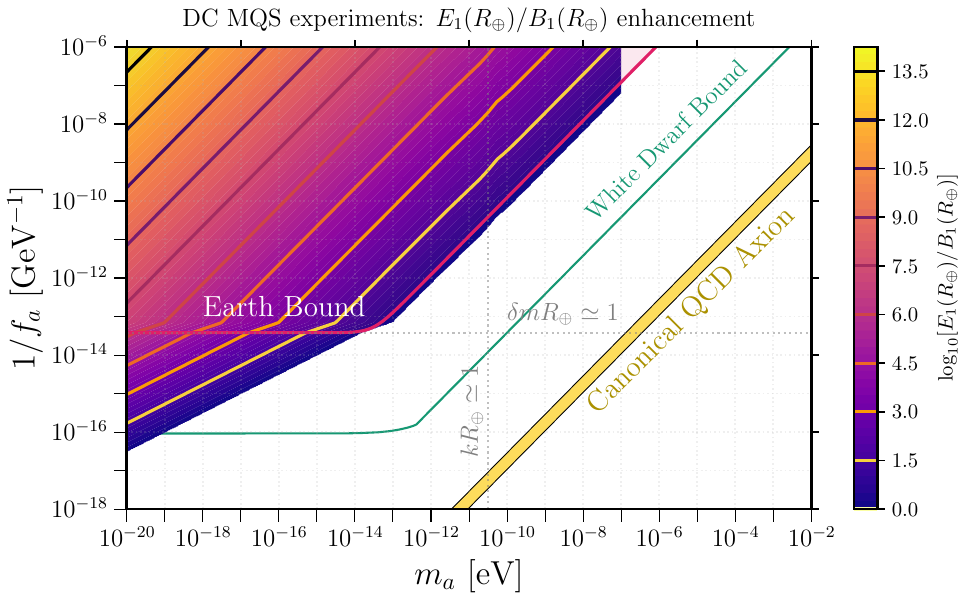}
    \caption{Region where the electric field signal is enhanced over the magnetic field signal in DC background field experiments operating in the magneto-quasistatic (MQS) limit. Most of the region with significant enhancement is contained within the region bounded by Earth-sourcing of axions~\cite{Hook:2017psm} and White Dwarf-sourcing \cite{Balkin:2022qer}. We restrict the shaded region to $m_a \lesssim 10^{-7}\,\text{eV}$, as above this mass, the MQS approximation breaks down for typical laboratory-sized experiments~\cite{Benabou:2022qpv}.}
    \label{fig:EfieldDCmqs}
\end{figure}

The result is that $E_1(R_\oplus) > B_1(R_\oplus)$ for sufficiently small values of $m_a,~f_a$, as shown in Fig.~\ref{fig:EfieldDCmqs}. Comparing with Fig.~\ref{fig:parspace}, we see that the induced electric field exceeds the induced magnetic field in the green region, where $\nabla a(R_\oplus)$ is affected, as well as in part of the blue region, where only $a(R_\oplus)$ is affected. The reason is that $B_1(R_\oplus)$ is sensitive to $a(R_\oplus)$, while $E_1(R_\oplus)$ is mostly sensitive to $\nabla a(R_\oplus)$, so if either of the relevant axion quantities is affected, a change in the ratio $E_1(R_\oplus)/B_1(R_\oplus)$ occurs. We discuss the detectability of signals further in Appendix~\ref{app:Vpt2}.

Not only is there a change in the ratio $E_1(R_\oplus)/B_1(R_\oplus)$, but we observe an important effect for experiments that are only sensitive to the magnetic field signal. 
The leading term in Eq.~\eqref{eq:B1mqs} scales as
\begin{align}
    B_1(R_\oplus) \propto \frac{C_\gamma \alpha}{2\pi f_a} a_0 \min\left[1, \frac{f_a}{(f_a)_\oplus^c}  \right]\ ,
    \label{eq:Bscaling}
\end{align}
where we ignore the effect of the resonances, which are sparsely populated for $kR \lesssim 1$.
Meanwhile, experiments set constraints on $C_\gamma/f_a$ assuming $B_1(R_\oplus) \propto (C_\gamma \alpha/2\pi f_a) a_0$. However, if $f_a< (f_a)^c_\oplus$, we observe that $B_1(R_\oplus)$ is \emph{independent of $f_a$}. Instead, they are actually sensitive to $C_\gamma a_0$, so a constraint on $g_{a\gamma\gamma}^0 a_0$ assuming vacuum axion \ac{DM} cannot be recast as a constraint on $a(R_\oplus)/f_a$. Instead, it leads to a constraint on $C_\gamma^\oplus/(f_a)^c_\oplus$. For sufficiently small values of $f_a \lesssim (f_a)^c_\oplus$, the new constraint on $C_\gamma^\oplus$ saturates the maximum value it can take in simple axion models. In Appendix~\ref{app:Vpt2}, we show a reinterpretation of $g_{a\gamma\gamma}^0$ in terms of $C_\gamma^\oplus$. This finding affects all experiments in the blue-green region of Fig.~\ref{fig:parspace}, and possibly extends into the pure blue region as well.

It should be noted, however, that most of the parameter space where the the electric and magnetic field signals are affected lies in the ``Earth Bound'' region. As with the axion gradient enhancement discussion above, the dynamics leading to this change in  \ac{EM} signals is the same as that which leads to the Earth Bound in the first place. Further, as discussed above, we have computed changes to the axion field assuming an expansion around a \ac{DM} vacuum. In the Earth Bound region, the true vacuum is shifted significantly, and our assumption of a \ac{DM} vacuum breaks down.

\subsubsection{Experiments with AC  EM fields}

These experiments are typically operated in the regime where $\w_0 l \gtrsim 1$
, where $\w_0$ is a resonant frequency of the apparatus
. This allows us to solve for the background fields  $\vect{E}_0^{\rm AC} \sim\vect{j}/\w_0,~ \vect{B}_0^{\rm AC} \sim \hat{\vect{n}}\times\vect{j}/\w_0$, where $\hat{\vect{n}}$ is a unit vector enforcing orthogonality of $\vect{E}_0,~\vect{B}_0$. 

In the regime where an AC  \ac{EM} field is present, we can solve Eqs.~\eqref{eq:E1eom} and~\eqref{eq:B1eom} in Fourier space, accounting for the fact that these experiments often operate in the resonant regime, so that we must include an Ohmic loss operator that depends on the device quality factor. The solution for $\vect{E}_1,~\vect{B}_1$ is obtained by decomposing each of these into a sum over eigenmodes, labeled by `$n$'. We find
\begin{align}
    \left(k_n^2 - i \frac{k_n \omega}{Q_n} - \omega^2 \right) \vect{E}_{1,n} \simeq - g_{a\gamma\gamma}(i \omega) E_0 \left( \hat{E}_0 \times \grad a_{\omega-\omega_0} - i m_a a_{\omega-\omega_0} (\hat{k}_0 \times \hat{E}_0)\right) \ , 
    \label{eq:E1heterodyne}
\end{align}
where $a_{\omega-\omega_0}$ indicates that the Fourier transform of the axion field should be evaluated at the frequency $\omega-\omega_0$. Evidently this will have the most support if $\omega-\omega_0 \sim m_a$ for \ac{DM} axions.
The quantity $\hat{E}_0$ is the orientation of the background $E$-field to zeroth order in $g_{a\gamma\gamma}$, $\hat{k}_0$ is the orientation of the background $E$-field propagation, and $Q_n$ is the quality factor of a given mode. This result assumes that the background  \ac{EM} field only has support at an angular frequency $\omega = \omega_0$, which is generally not the case due to oscillator phase noise~\cite{Berlin:2019ahk,Berlin:2020vrk}. 

From Eq.~\eqref{eq:E1heterodyne} we can see that the axion gradient contribution (first RHS term) will dominate over the usual time-derivative contribution (second RHS term) in the same region where $\vect{E}_1$ dominates over $\vect{B}_1$ for a DC MQS experiment. The reason is that heterodyne experiments operate in the radiation regime where $\vect{E}_{0,1}\sim \vect{B}_{0,1}$. Therefore, the comparison in the source terms for both $\vect{E}_1$ and $\vect{B}_1$ is made between $\grad a(R_\oplus)$ and $\partial_t a(R_\oplus)$, as can be seen in Eqs.~\eqref{eq:E1eom},~\eqref{eq:B1eom}. This is the same comparison we found for DC MQS experiments with zero background electric field, except with $\grad a(R_\oplus)$ and $\partial_t a(R_\oplus)$ as sources for $\vect{E}_1$ and $\vect{B}_1$ respectively.  

There is another key difference between heterodyne and MQS experiments, which is the readout. Experiments in the MQS limit are usually $B$-field optimised, and are therefore not necessarily sensitive to the increased $E$-field due to the enhanced axion gradient. Heterodyne experiments operate in a regime where they can be equally sensitive to $E$ or $B$ fields. However, it should be noted that the axion gradient-enhanced sensitivity will only exist as long as the overlap between the signal mode $\vect{E}_{1,n}$ and $\hat{E}_0 \times \hat{r}_\oplus$ is non-zero. The unit vector $\hat{r}_\oplus$ points in the radial direction in a coordinate system centered on Earth, while $\hat{E}_0$ is in an unspecified coordinate system.

Full evaluation of the possible enhancement of low-mass axion signals for heterodyne experiments is beyond the scope of this paper. As in the case of DC MQS experiments above, $E_{1}$ for heterodyne experiments is proportional to the same combination as $B_1$ in Eq.~\eqref{eq:Bscaling} above, so our statements regarding reinterpretation of $g_{a\gamma\gamma}^0$ in terms of $C_\gamma^\oplus$ hold if $\hat{E}_0 \times \hat{r}_\oplus = 0$. Meanwhile, if $\hat{E}_0 \times \hat{r}_\oplus \neq 0$, the enhanced gradient contribution to $E_1$ affects the expected sensitivity. We show some results in Appendix~\ref{app:Vpt2}.

Unfortunately, no existing AC  \ac{EM} field experiment targets this parameter space, so there are no datasets to be re-analysed in light of our findings. However, proposed AC  \ac{EM} field experiments can in principle target the relevant parameter space, and a dedicated analysis of the impact of our findings for each one should be conducted.

\section{Conclusions}\label{sec:conclusions}

In this work, we consider a class of ultralight, sub-eV, spin-0 \ac{DM} models with quadratic couplings to matter. It consists of a number of well motivated models (see~\cite{Banerjee:2022sqg} for a recent discussion).
Particularly interesting is the case of the QCD axion, which has an unavoidable quadratic interaction with nuclear matter, arising from its defining coupling $(a/f_a)G\tilde{G}$. 
In this work, we studied the intricate effects that Earth's presence has on the axion \ac{DM} field. A key element of our analysis is the inclusion of finite momentum for the \ac{DM} solution far away from the source. We have assumed that the axion \ac{DM} is virialized, and thus its momentum follows the Maxwell-Boltzmann distribution. 
We have not included in our treatment the presence of gravitational interactions that could lead to focusing~\cite{Kim:2021yyo} or a formation of Solar/Earth halos~\cite{Budker:2023sex,Banerjee:2019epw}, which might significantly alter the kinematic distribution of \ac{DM}. 
We also de facto approximated the velocity distribution of axions to be spherically symmetric; however, as discussed in Appendix~\ref{app:pert}, this may wash away some non-negligible effects. 

\begin{figure}[t]
    \centering
    \includegraphics[width=0.8\linewidth]{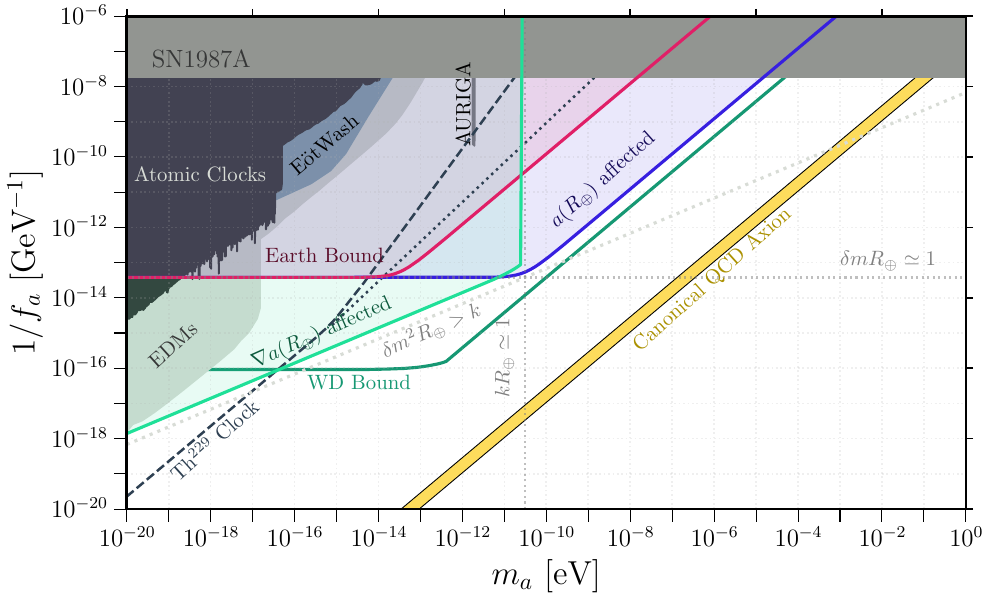}
    \caption{We show the regions affected by Earth as in Fig.~\ref{fig:parspace}, but additionally superimposed on a set of Earth-based experimental constraints (in shades of grey) on axions from CP-even operators~\cite{Wagner:2012ui,Branca:2016rez,Abel:2017rtm,Roussy:2020ily,Schulthess:2022pbp,JEDI:2022hxa,Zhang:2022ewz,Beadle:2023flm, Kim:2023pvt,Madge:2024aot}. 
    Additionally, we show the bound from SN1987A \cite{Springmann:2024ret} (dark grey shaded region) and from white dwarfs \cite{Balkin:2022qer} (region above green line). 
    We also show the projected sensitivity of a future $^{229}\text{Th}$ nuclear clock (dark grey dashed), including the sensitivity to stochastic effects (dark grey dotted line)~\cite{Caputo:2024doz,Fuchs:2024edo}. }
    \label{fig:summaryplot}
\end{figure}

Our main findings are that for axion masses roughly above $10^{-14}\,$eV (in the blue region of Fig.~\ref{fig:summaryplot}), the amplitude of the axion field on Earth could be significantly suppressed relative to the canonical value of $\langle a(t)^2\rangle= \rhoDM/m_a^2$. We find that typically $a(R_\oplus) \leq a_0$ in the blue region of Fig.~\ref{fig:summaryplot}, with important consequences for experiments conducted on Earth.
Furthermore, for masses below roughly $10^{-11}\,$eV (the pure green region of Fig.~\ref{fig:summaryplot}), the radial component of the spatial gradient of the axion field can be greatly enhanced compared to its canonical value. 

Our findings have important implications for theories where \ac{DM} contains both quadratic and linear interactions. This is the case of the celebrated QCD axion, but also other axion-like theories with moderate CP violation, such as some relaxion-models~\cite{Banerjee:2018xmn}.
For instance, experiments searching for low-mass axions via the photon interaction in a DC background field (e.g.,~\cite{DMRadio:2022jfv,DMRadio:2022pkf,DMRadio:2023igr}) may benefit from measuring the axion-induced electric field, which is enhanced. Finally, experiments operating with comparable background electric and magnetic fields, such as heterodyne detectors~\cite{Berlin:2019ahk,Berlin:2020vrk,Giaccone:2022pke}, can also benefit from the enhancement of the axion DM gradient. 
One particularly exciting possibility would be, for instance, in the case where a signal would appear both in magnetic-gradient-based experiments, sensitive to the linear coupling, as well as in nuclear clock comparisons, which are sensitive to the quadratic coupling. The comparison between the two would enable us to study directly the above matter effects.

Let us briefly discuss a few limitations of this work, and potential directions for future studies. 
Within this work, we adopted a framework where Earth is spherically symmetric and perfectly uniform. 
This simplification is valid when the axion's momentum $k$ and the Earth-induced axion mass reduction $\dm$ are small compared to $1/R_{\oplus}$, but unreliable beyond it. 
Imperfections of Earth are likely to broaden, shift, mix, or otherwise modify some of the resonances found in this work. For example, for $kR_{\oplus}\ll 1, \dm R_{\oplus}\gg 1$ we ignore resonances due to their sparseness, but should resonances be broadened, it is not obvious their contribution is negligible. In addition, we did not account for the presence of other nearby celestial bodies (such as the Sun, Jupiter, or Moon). 
We have also assumed that the system is static and focused on the steady-state solution. Thus, we completely ignored questions associated with what happens during the formation of Earth (or the Solar System) and how the \ac{DM} field reaches its asymptotic value.

One highly relevant question which is not fully addressed by this work is whether or not canonical QCD axion models (e.g., the KSVZ~\cite{Kim:1979if,Shifman:1979if} and DFSZ~\cite{Dine:1981rt,Zhitnitsky:1980tq} models) are affected by this effect. 
Within our calculations, we find that for such axions with a velocity close to $10^{-3}$, the correction to their momentum is much smaller than their vacuum-momentum. And yet, treating the potential perturbatively (see Appendix~\ref{app:pert}) tells us the field phase is non-negligibly modified by the presence of Earth, and it is therefore conceivable that some effects are present (especially for an experiment located at the antipodal point of the \ac{DM} average incoming velocity). Alternatively, should non-virially distributed \ac{DM} be present, especially if the average momentum and its standard deviation are reduced, a larger fraction of the parameter space is affected. 
We leave a dedicated exploration of this to future work.

Finally, while we have focused on terrestrial experiments and observables, it is interesting to consider other celestial bodies. For example, even for canonical QCD axion models, major changes in the field amplitude would be present in the vicinity of neutron stars, which may be important for bounds relying on the conversion of axions to photons in the magnetosphere of a neutron star~\cite{Hook:2018iia}. 
Whether or not this produces observable effects for white dwarfs is also an interesting question. As mentioned before, the presence of the Sun or Jupiter may also introduce interesting effects which may motivate sending highly accurate sensors to either be closer or further away from them. 

In our analysis, we have given a careful assessment of matter effects on spin-0 wave-like dark matter, accounting for its momentum distribution in the process. Our results demonstrate that great care must be taken in such an analysis, as the results are sensitive to the EFT being used, the vacuum around which one expands, and the velocity distribution of dark matter. Nevertheless, we have demonstrated that matter and momentum both matter for correctly determining the optimal experimental strategy to search for dark matter on Earth.

\section*{Acknowledgments}

During the final stages of preparation, we became aware of similar work by the authors of Ref.~\cite{Joerg}. We thank them for coordinating the release of our works.

We thank Martin Bauer, Kai Bartnick, Chelsea Bartram, Salvatore Bottaro, Sreemanti Chakraborti, Michael Geller, Anson Hook, Simon Knapen, Ben Meiring, Hitoshi Murayama, Nadav Outmezguine, Nick Rodd, Stefan Stelzl, Ofri Telem, Andreas Weiler, Linda Weishuang Xu, and the Berkeley BCTP group for discussions. 
The work of AB is supported by the
National Science Foundation under grant number
PHY2210361 and the Maryland Center for Fundamental Physics.
The work of SARE was supported by SNF Ambizione grant PZ00P2\_193322, \textit{New frontiers from sub-eV to super-TeV}.
The work of Y.~V.~S.~was supported by the Australian Research Council under the Discovery Early Career Researcher Award No.~DE210101593.
The work of GP is supported by grants from NSF-BSF, ISF and Minerva. 
AB, GP, and YVS thank Joshua Eby for early collaboration and many fruitful discussions. 
The same authors are also grateful to the Munich Institute for Astro-, Particle and BioPhysics (MIAPbP)  where part of the work was done.

\appendix

\section{Virial Averaging and the Signal Power Spectral Density}\label{app:virial}

For a virialized \ac{DM} halo, let us find the expectation value for the field's power. 
We find the virial averaged squared field amplitude is 
\begin{align}
    \langle a^2 \rangle = 2\int d^3k \abs{a_{\omega}}^2 f_k\approx 2\left(\sqrt{\pi}\kvir\right)^{-3}\int d^3k \abs{a_{\omega}}^2 e^{-\frac{\abs
{\vec{k} - \vec{k_\oplus}}^2}{ \kvir^2}}\,,
\end{align}
 with the normalization of $\int d^3k f_k=1$, and $k_{ \oplus}\equiv m_a v_{\oplus}\approx 10^{-3}m_a$ is the galactic velocity of the solar system. 
The integral reads 

\begin{align}
    \tfrac{1}{2}\langle a^2 \rangle & \approx \left(\sqrt{\pi}\kvir\right)^{-3}\int d^3k \abs{a_{\omega}}^2 e^{-\frac{\leri
{\abs{\vec{k}}^2 -2\abs{k}\abs{k_\oplus}\hat k\cdot\hat k_{\oplus} +\abs{\vec{k_\oplus}}^2}}{ \kvir^2}} \nonumber\\
&=2\pi^{-1/2}\kvir^{-3}\,\sum_{\ell,\ell'}\int \, dk\,k^2\, a_{\omega,\ell}a^*_{\omega,\ell'} 
     e^{-\frac{\leri{k^2+k_\oplus^2}}{\kvir^2}}\int \frac{d\Omega}{2\pi}\,
P_\ell(\hat{r}\cdot\hat{k})\,
P_{\ell'}(\hat{r}\cdot\hat{k})\, 
e^{\frac{{
2\abs{k}
\abs{k_\oplus}
\hat{k}\cdot\hat{k}_{\oplus}
}}
{\kvir^2}}\,\,.
\end{align}

Looking at the angular part only, we get

\begin{align}
    \int_{-1}^1 dx\, P_\ell(x) P_{\ell'}(x) e^{\frac{2 k k_{\oplus}x \cos {\theta }}{\kvir^2}} I_0\left(\frac{2 k k_{\oplus} \sqrt{1-x^2} \sin {\theta}}{\kvir^2}\right) \,\,,
\end{align}
where $x=\hat{r}\cdot\hat{k}$ and $\sin\theta$ is the angle associated to $\hat{k}_{\oplus}$. 
We see that modes where $\ell$ is far from $\ell'$ are suppressed when $k\sim k_{\rm vir}$. 

For simplicity, in most of this work, we focus only on $\ell=\ell'$ terms, essentially taking $k_{\oplus}\to 0$. Another way to reach a similar result is averaging over $\hat{k}_{\oplus}\cdot {\hat r}$, essentially asking what is the average observable for a randomly distributed location on Earth. As demonstrated in appendix~\ref{app:pert}, this simplification can lead to an artificial suppression of Earth's effect.

Furthermore, should a measurement last longer than $1/|\omega_k-\omega_{k'}|$, the measured power spectral density of the field would allow distinguishing $k$ and $k'$. We also neglect this effect, essentially treating measurements as if they have no frequency resolution and cannot determine the lineshape of the axion.

\section{Convergence Proofs}
\label{app:conv}

In this section, we show the convergence of the $\ell$ series that we use in Eq.~\eqref{eq:oursolout} and~\eqref{eq:oursolin} for $k>0$. 
We start our discussion by showing that none of the individual $\ell$ terms diverges for $k>0$, and if $\ell>1$, even the $k=0$ limit is well defined, in Appendix~\ref{app:conv1}. 
In Appendix~\ref{app:conv2}, we show that their infinite sum is convergent as well. 
We conclude the section by showing that the series is still convergent even after integrating over the thermal distribution (which includes a $k\to 0$ contribution) in Appendix~\ref{app:conv3}. 

\subsection{Individual coefficients non-divergence}\label{app:conv1}

In the main text, we divide the solution to the field EOM into two parts: solution outside the dense object (Eq.~\eqref{eq:oursolout}), and inside the dense object (Eq.~\eqref{eq:oursolin}). 
By matching them at the surface of the source, we obtain a continuous smooth solution over all $r$. 
We consider a spherically symmetric source, and thus the solutions are expressed in terms of $m=0$ spherical harmonic as $a_{\omega}^{\rm in \,(out)}(r,\theta) = \sum_{\ell=0}^{\infty} a_{\omega,\ell}^{\rm in \,(out)}(r,\theta)$\,, where $a_{\omega, \ell}^{\rm in \,(out)}$ are given in Eq.~\eqref{eq:al}. 
Notice $a_{\omega, \ell}$s are given by the ratios of the spherical Bessel Functions, $j_\ell$ and/or $y_\ell$, of different coefficients, and can written in the form of $a_{\omega, \ell}\propto f(j_\ell,y_\ell)/g(j_\ell,y_\ell)$ where $f$ and $g$ are some generic functions. 
Thus, if $f(j_\ell,y_\ell)$ does not diverge and $g(j_\ell,y_\ell)$ has no zeros for any $\ell$, then $a_{\omega, \ell}$ is finite. 

First, let us discuss the numerators. 
By construction, the solution inside the source is regular as $r\to 0$. 
On the other hand, as $\{j_\ell(r),y_\ell(r)\}\to 1/r \times (\cos(r),\sin(r))$ as $r\to \infty$, the numerator is regular for solution outside the source as well. 
Note that, as $y_\ell(r)\propto r^{-(\ell+1)}$ as $r\to 0$, there is an apparent divergence of $a^{\rm out}_{\omega}$. 
However as $r\to 0$, one should use the interior solution, so this divergence is never relevant. 
Therefore, all we now need to show is that the denominators for both the solutions are always non-zero for $k>0$.

First we notice that the denominators of the inside and the outside solutions differ by a multiplicative factor of $k$. 
As we are considering the case of $k\neq 0$, the denominator of interest for both the solutions is $Q_\ell$, which we write explicitly as

\begin{equation}
Q_\ell=\Big(kj_{\ell}(\kd R)y_{\ell+1}(kR)-\kd j_{\ell+1}(\kd R)y_\ell(kR)\Big)+i\Big(\kd j_\ell(kR)j_{\ell+1}(\kd R)-kj_\ell(\kd R)j_{\ell+1}(kR)\Big)\,.
\end{equation}

For $Q_\ell$ to vanish, both its real and imaginary parts must simultaneously be zero. 
We will start with the assumption that this condition is met, only to demonstrate that it leads to a contradiction. 

Before we start the proof, let us mention a lemma~\cite{besselcite}: if $J_\nu(x)=0$ and $x>0$ and $\nu \in \mathbb{R}$, then $J_{\nu+m}(x)\neq 0$ for all non-zero integers $m$. Since $j_\ell\propto J_{\ell+1/2},y_m\propto J_{-m-1/2}$ for positive arguments and integer orders, all $y$s and $j$s have no common zeros with each other or themselves.

Now, as $\ell$ is an integer, using the above lemma we obtain, for ${\rm Re}(Q_\ell)$ to vanish if $y_{\ell+1}(kR)=0$, then $j_{\ell+1}(\kd R)$ will have to vanish and vice-versa.  
This leads to ${\rm Im}(Q_\ell)= -kj_{\ell}(\kd R)j_{\ell+1}(kR)\neq 0 $ using the above lemma. 
Thus when $Q_\ell=0$, $y_{\ell+1}(kR)\neq 0$.
For non-zero $y_{\ell+1}(kR)$, one can take $j_{\ell}(\kd R)=\tfrac{\kd }{k}j_{\ell+1}(\kd R)y_\ell(kR)/y_{\ell+1}(kR)$ for ${\rm Re}(Q_\ell)$ to vanish. 
With that choice, we obtain ${\rm Im}(Q_\ell)= -\kd j_{\ell+1}(\kd R)/(k^2R^2y_{\ell+1}(kR))$  
using the Bessel function identity of $j_\ell(x)y_{\ell+1}(x)-j_{\ell+1}(x)y_{\ell+1}(x)=-1/x^2$. 
Thus, ${\rm Im}(Q_\ell)=0$ requires $j_{\ell+1}(\kd R)=0$ which in turns contradicts the lemma as $j_{\ell+1}(\kd R)\propto j_\ell(\kd R)\neq 0$ (obtained from ${\rm Re}(Q_\ell)=0$). 
Thus, we have reached a contradiction. 
Therefore, no individual term, $a^{\rm in\,, out}_{\omega,\ell}(r,\theta)$, can be divergent, for $k>0$.  
\\

Let us now show the behavior for $k\to 0$. 
Note that in that case $q=\sqrt{k^2+\dm^2}\simeq \dm$. 
By expanding in small powers of $k$, we obtain

\begin{equation}
\!\!\!
{\rm Re}(Q_\ell)=-\frac{(kR)^{-1-\ell}(2\ell)!/\ell!}{2^{\ell+1}(2\ell-1)\dm R^2 }\Bigg[(4\ell-2)(\dm R)^2 j_{\ell-1}(\dm R)-(1+2\ell)(kR)^2(\dm R)j_\ell(\dm R)\Bigg]+\mathcal{O}\Big((kR)^{3-\ell},(kR)^{1-\ell}\Big)\,,
\end{equation}

and
\begin{equation}
{\rm Im}(Q_\ell)=-\frac{2^{\ell}(kR)^{\ell}j_\ell(\dm R)\, \ell!}{R\, (2\ell)!}+\mathcal{O}\Big((kR)^{\ell+2}\Big)\,.
\end{equation}

Since $j_\ell,j_{\ell-1}$ do not share zeroes for finite arguments, ${\rm Re}(Q_\ell)$ scales as $(kR)^{-1-\ell}$ or $(kR)^{-\ell+1}$. 
As, ${\rm Im}(Q_\ell)\propto (kR)^\ell$, ${\rm Re}(Q_\ell)$ dominates over the imaginary part of $Q_\ell$ for $\ell\geq 2$ in the $k\to 0$ limit. 
Thus, $a^{\rm in}_{\omega,\ell}(r,\theta)\propto 1/(kR\, Q_\ell)$ scales as either  $(kR)^{\ell}$ or $(kR)^{\ell-2}$, both of which are finite.

Note that, $a^{\rm out}_{\omega,\ell}$ has terms in the form of $ y_\ell(kr) {\rm Im}(Q_\ell)/Q_\ell$ and $j_\ell(kr) {\rm Re}(Q_\ell)/Q_\ell$. 
Generally, for $\ell\geq 2$, ${\rm Re}(Q_\ell)$ dominates over ${\rm Im}(Q_\ell)$, and $j_\ell$ is converges to 0. 
Therefore, in $a^{\rm out}_{\omega,\ell}$, the term proportional to ${\rm Re}(Q_\ell)j_{\ell}(kR)/Q_\ell$ is finite. 
Similarly, as $y_\ell(kr)\propto (kr)^{-\ell-1}$, the term proportional to ${\rm Im}(Q_\ell)y_{\ell}(kr)/Q_\ell$ scales as $(kR)^{\ell}$ or $(kR)^{\ell-2}$, leading to a finite contribution. 
Conversely, either by applying the above scaling, or using the explicit expressions given in Eq.~\eqref{eq:a0} and~\eqref{eq:a1}, we see that both $\ell=0,1$ scale as $1/(kR)$ on resonances. 
This scaling is correct for the solution both inside and outside the dense object, as well as for the gradients.

\subsection{Infinite series convergence}\label{app:conv2}

After proving the convergence of each $a^{\rm in\,, out}_{\omega,\ell}(r,\theta)$, we want to prove their infinite sum over $\ell$ i.e. $a_{\omega}^{\rm in \,,out}(r,\theta) = \sum_{\ell=0}^{\infty} a_{\omega,\ell}^{\rm in \,,out}(r,\theta)$ is convergent as well. 
As none of the $\ell$ terms are divergent for $k>0$, we may remove any finite number of them, and the new series converges if and only if the full series converges.
\\

For a fixed $k,\kd ,R,r$, let us show that the sum absolutely converges i.e. $\sum_{\ell=0}^{\infty} |a_{\omega,\ell}^{\rm in \,,out}(r,\theta)|$ is convergent. 
An absolute convergence,  $\sum_{\ell=0}^{\infty} |a_{\omega,\ell}^{\rm in \,,out}(r,\theta)|$ implies convergence of $\sum_{\ell=0}^{\infty} a_{\omega,\ell}^{\rm in \,,out}(r,\theta)$ in Banach space (see {\it e.g.}~\cite{folland_real_analysis}). 
Let us assume we start our sum at some $\ell=L>0$, and as we examine the terms, we will find what $L$ needs to be. 
Therefore, we examine the following sum  
\begin{equation}
\Sigma_1=\sum_{\ell=L}^{\infty}|P_\ell(\cos(\theta))c_\ell(k,\kd ,r,R)|\,,
\end{equation}
with $c_\ell(k,\kd,r,R)$ is either $a^{\rm out}_{\omega,\ell}(r,\theta)/P_\ell(\cos(\theta))$ or $a^{\rm in}_{\omega,\ell}(r,\theta)/P_\ell(\cos(\theta))$ for $r\geq R$ and $r<R$ respectively. 
Since $|P_\ell(\cos(\theta))|\leq 1$, $\Sigma_1$ absolutely converges if 
$\Sigma_2=\sum_{\ell=L}^{\infty}|c_\ell(k,\kd ,r,R)|$
converges. 
\\

We will begin by proving convergence of $\Sigma_2$ inside Earth i.e. $r<R$. For convenience, we define ${\tilde j_\ell},\,{\tilde y_\ell}$ as,

\begin{equation}
j_\ell(x)=\frac{\sqrt{\pi}}{2\Gamma(3/2+\ell)}\left(\tfrac{x}{2}\right)^\ell \tilde j_\ell(x)\,,\,\,\, 
y_\ell(x)=-\frac{\Gamma(1/2+\ell)}{2\sqrt{\pi}}\left(\tfrac{x}{2}\right)^{-\ell-1} \tilde y_\ell(x)\,.
\end{equation}

Now we will proceed with the convergence of $\Sigma_2$ proof with the assumption that asymptotically $\tilde j_\ell,\,\tilde y_\ell$ approach unity i.e. for $\ell\to \infty$, $\tilde j_\ell,\,\tilde y_\ell\to 1$ (\textit{lemma (i)}). 
We will prove \textit{lemma (i)} later.

As $c_\ell=a_{\omega,\ell}/P_\ell(\cos(\theta))$, like earlier we can write it as the ratio of a numerator and a denominator. 
For $a^{\rm in}_{\omega,\ell}$, the denominator is $kR^2\,Q_\ell$. 
To show absolute convergence, we may always decrease from the denominator. 
By removing the imaginary part of $Q_\ell$, we get,

\begin{equation}\label{eq:denomlargelpt1}
|Q_\ell|\geq \left|{\rm Re}(Q_\ell)\right|=
\tfrac{1}{kR^2}\left(\tfrac{\kd }{k}\right)^{\ell}\Big|\tfrac{(\kd R)^2}{3+8\ell+4\ell^2}\tilde j_{\ell+1}(\kd R)\,\tilde{y}_\ell(kR)-\tilde{j}_\ell(\kd R)\, \tilde{y}_{\ell+1}(kR)\Big|\,.
\end{equation}

We will later show in {\it lemma (i)} that both $\tilde j_\ell,\tilde y_\ell$ approach 1 for $\ell\to \infty$. 

Assuming that, we take $L$ such that, for any $\ell>L$, $1/2 \leq\tilde j_\ell,\tilde y_\ell\leq 2$. 
With this choice of $L$, the second term in the absolute value in Eq.~\eqref{eq:denomlargelpt1} is at least $1/4$. 
Furthermore, by taking $L>4(\kd R)$, which is to make sure there is no accidental cancellation between the first and the second term, we get the first term would be at most $4/(4\cdot 4^2)=1/16$. 
Since $|a-b|\geq ||a|-|b||\geq ||a|_{\rm min}-|b|_{\rm max}|$ for $|a|>|b|$, we obtain,

\begin{equation}
kR^2\,|Q_\ell|\geq \left(\tfrac{\kd }{k}\right)^{\ell}\left|\frac{1}{16}-\frac{1}{4}\right|=\tfrac{3}{16}\left(\tfrac{\kd }{k}\right)^{\ell}.
\end{equation}

For the inside solution, the  numerator is  $|i^{\ell}(1+2\ell)j_\ell(\kd r)| $.
Combining  \textit{lemma (i)}, and the choice of $L$, we find that the term in the sum is 
\begin{equation}\label{eq:boundin}
|c_\ell(k,\kd ,r,R)|_{r<R}\leq \frac{2^{5-\ell}(kr)^\ell\sqrt{\pi}}{3\Gamma(1/2+\ell)},
\end{equation}

which converges for all $k,r$. Note that we have implicitly assumed that $r>0$; though for $r=0$ only $\ell=0$ is non-zero, and thus $c_\ell$ is regular.

We now need to show that the series converges also for the outside solution i.e. for $r\geq R$ as well. 
While we can use the same lower bound for $Q_\ell$, we need to calculate the upper bound for the numerator. 
It includes the summation of different $\tilde y$s and $\tilde j$s, and some of them are with relative signs to each other.  
However, using triangle inequality,  $|a_1+...+a_n|\leq |a_1|+...+|a_n|$, we simply sum the absolute values of individual terms and find an upper of that.  
Like before, we choose an appropriate $L$ so that each ${\tilde j}_\ell,{\tilde y}_\ell\leq 2$ for any $\ell>L$.  
This gives us the following upper bound

\begin{equation}\label{eq:boundout}
|c_\ell(k,\kd ,r,R)|_{r\geq R}\leq \frac{2^{5-\ell}k^lr^{-\ell-1}\sqrt{\pi}\left((k^2+\kd ^2)R^{3+2\ell}+r^{1+2\ell}(3+4\ell(2+\ell)+(\kd  R)^2)\right)}{3\Gamma(5/2+\ell)}\,,
\end{equation}
which once again has an infinite convergence radius, due to the $1/\Gamma(5/2+\ell)$ in large $\ell$ limit. 
\\

Now we are left with proving our \textit{lemma (i)} which is to show that for $\ell\to \infty$, $\tilde j_\ell,\,\tilde y_\ell\to 1$. 
Let there be $\epsilon>0$. We need to show that there exists an $L$, for which for any $\ell>L$, $|\tilde{j}_\ell(x)-1|<\epsilon $ and $|\tilde{y}_\ell(x)-1|<\epsilon$. 
Note that, for us $x>0$, and it only asymptotically approaches to $0$. 
Let us start by proving this for $\tilde{j}_\ell$.
Starting from the infinite series expansion for the Spherical Bessel functions of the first kind $j_{\ell}(x)$ which has infinite convergence radius, and using the definition of $\tilde{j}_\ell$, we obtain,  

\begin{equation}
|\tilde j_\ell(x)-1|=\Bigg|\sum_{k=1}^\infty\frac{\Gamma(\frac{3}{2}+\ell)(-1)^k(\frac{x}{2})^{2k}}{k!\, \Gamma(\frac{3}{2}+\ell+k)}\Bigg|\leq \frac{\Gamma(\frac{3}{2}+\ell)}{\Gamma(\frac{3}{2}+\ell+1)} \sum_{k=1}^\infty\frac{(\frac{x}{
2})^{2k}}{k!}=\frac{2\, e^{x^2/4}}{3+2\ell}<\frac{e^{x^2/4}}{L}\,.
\end{equation}

Thus by choosing $L>e^{x^2/4}/\epsilon$, we find that indeed $|\tilde j_\ell(x)-1|<\epsilon$. 
Note that in practice, already for $\ell\sim x$, the approximation ($\tilde j_\ell(x)\sim 1$) works, so usually one only needs to examine $\ell$s up to $\kd R$ in relation to the formulas discussed in the main text. 
Similarly, to prove {\textit{lemma (i)}} for $\tilde{y}_\ell$, we use similar series expansion and obtain 

\begin{equation}
|\tilde{y}_\ell(x)-1|=\Bigg|\sum_{k=1}^\infty\frac{\Gamma(\frac{1}{2}-k+\ell)(\frac{x}{2})^{2k}}{k!\, \Gamma(\frac{1}{2}+\ell)}\Bigg| \leq \frac{\Gamma(\ell-\frac{1}{2})}{\Gamma(\ell+\frac{1}{2})}
\sum_{k=1}^\infty\frac{(\frac{x}{2})^{2k}}{k!}=\frac{2\, e^{x^2/4}}{2\ell-1}<\frac{e^{x^2/4}}{L}\,.
\end{equation}

Thus, by choosing $L>e^{x^2/4}/\epsilon$, we obtain $|\tilde y_\ell(x)-1|<\epsilon$. 
Note that in the above proof we also that for any integer $n$, $(-1)^n\Gamma(-n+1/2)\Gamma(n+1/2)=\pi$. 

\subsection{Convergence of the virally averaged integral}\label{app:conv3}

In this subsection, we address the finite nature of the virial  averaged integrals i.e.
$\int_{0}^{\infty} dk\, k^2 |a_{\omega,\ell}|^2$ is finite even for arbitrarily small and arbitrarily large $\kvir$. 
As the standard halo model of \ac{DM}, does not have arbitrarily large momenta, we only discuss issues related to the $k\to 0$ limit.

We have shown in previous two subsections that for a small enough $k$, all $\ell\geq 2$ modes are bounded by non-negative powers of $k$. 
Hence, their value for $k\to 0$ is easily analytically continued, and they are within $L_2$ space. 
On the other hand, the $\ell=0,1$ modes diverge at most as $k^{-1}$. 
Thus the virial averaged integrals scale as 
$\propto \int d^3k\,\, 1/k^2\propto k$, and gives a finite result.

\section{Resonant Contributions}\label{app:resonances}

In this appendix, we clarify the intention behind the terminology of ``resonances'', which correspond to the observed drastic enhancements of the axion field value within a narrow band in momentum space (see Fig.~\ref{fig:l3sol} for instance).

To begin with, it is useful to invoke the asymptotic behavior of spherical Bessel functions to understand where and how often resonances will occur.
In order to obtain a sharp resonance, it is necessary that \textit{i)} ${\rm Re}(Q_\ell)$ and ${\rm Im}(Q_\ell)$ take different typical values within some region of $k$ values and that \textit{ii)}, within that $k$-region, the hierarchy between ${\rm Re}(Q_\ell)$ and ${\rm Im}(Q_\ell)$ is flipped in a small $k$-subregion (or at least in isolated small $k$-subregions). 
The primary difference between ${\rm Re}(Q_\ell)$ and ${\rm Im}(Q_\ell)$ is that the former depends on $y_{\ell}(kR),y_{\ell+1}(kR)$, while the latter depends on $j_\ell(kR),j_{\ell+1}(kR)$. 
The region wherein these functions take different typical values is where $kR\lesssim\lt$, with $\lt\equiv \ell+1/2$, where generically ${\rm Re}(Q_\ell)$ is much bigger than ${\rm Im}(Q_\ell)$. A resonance is thus found in the (isolated) sub-regions of $kR\lesssim\lt$ where the hierarchy between ${\rm Re}(Q_\ell)$ and ${\rm Im}(Q_\ell)$ is flipped. Given its functional form, this really means that ${\rm Re}(Q_\ell)$ vanishes somewhere in these subregions. That this can happen imposes a further constraint on $\kd R$. Indeed, when $\kd R\lesssim\lt$, $\abs{j_\ell(\kd R)}$ decrease with $\ell$, while the opposite holds for $\abs{y_\ell(\kd R)}$. Furthermore, ${\rm Re}(Q_\ell)=0\iff j_\ell(\kd R)y_{\ell+1}(k R)=j_{\ell+1}(\kd R)y_\ell(k R)$. Therefore, $kR,\kd R\lesssim \lt$ would make the left-handside much larger than the right-hanside, and the equality could not be fulfilled. Putting all this together, we find that the resonances only occur for $kR\lesssim \lt\lesssim \kd R$ (or possibly for the subregion of $kR\ll \lt \ll \kd R$).

We thus now define $\kres$ as a momentum for which ${\rm Re}(Q_\ell)=0$. 
In principle, there could be multiple such $\kres$ values, for a given $\ell$. 
Furthermore, we define the associated $\kdres$ as the momentum inside Earth for this $\kres$ value. 
Our goal is to estimate the size of the field, for $k=\kres+{\dk}$, where ${\dk}$ is small enough to fulfill a condition that will be discussed later. Expanding ${\rm Re}(Q_\ell)$ and ${\rm Im}(Q_\ell)$ at leading order in $\delta k$ in the region $kR\lesssim \lt\lesssim \kd R$, we find 
\begin{equation}
    Q_\ell\approx-\frac{\kres  R{\dm}^2 j_\ell(\kdres R)y_{\ell-1}(\kres R) y_{\ell+1}(\kres R)}{\kdres^2 y_\ell(\kres R)}{\dk}-i\frac{j_{\ell}(\kdres R)}{y_\ell({\kres R})\kres R^2},
\end{equation}
where we used several Bessel identities, and the fact that ${\rm Re}(Q_\ell)|_{k=\kres}=0$. From the above, we can identify the width of the resonance, $\frac{2 \kdres^2}{\kres^2 R^3 \delta m^2 \abs{
  y_{\ell-1}(\kres R) y_{\ell+1}( \kres R)}}$, which scales like $\ell^{-2\ell}\(\kres R\)^{2\ell}$.

Using the above, we can estimate the contribution of a given resonance,
\begin{equation}
\frac{1}{4\pi}\int d\Omega_k \int d{\dk}\,  |a_{\omega,\ell}(R,\theta)|^2=\frac{a_0^2}{R}\left|\frac{(\kdres R)^2(1+2\ell)\pi y_\ell(\kres R)^2}{(\kres R)^2({\dm} R)^2y_{\ell-1}(\kres R)y_{\ell+1}(\kres R)}\right|\,,
\end{equation}
where we took the integral over $d{\dk}$ to be from $-\infty$ to $\infty$, since resonant contribution is dominant, and the regions far from it contribute negligibly in the lorentzian approximation.

Within the regime of interest, the ratio $\tfrac{y_{\ell}(kR)^2}{y_{\ell-1}(kR)y_{\ell+1}(kR)}$ 
is roughly one. 
Therefore, if we take the Boltzmann distribution of axion momenta to be roughly $\pi^{-3/2}k_{\rm vir}^{-3}e^{-(|\mathbf{k}|/k_{\rm vir})^2}$, we get
\begin{equation}
\left.\int d^3k f_{\rm MB}(k)|a(r,\theta)|^2\right|_{\rm resonances}=\sum_{\ell,\kres}\frac{8\sqrt{\pi}a_0^2e^{-\kres^2/k_{\rm vir}^2}\kdres^2 (1+2\ell)}{(k_{\rm vir} R)^3{\dm}^2} \ .
\end{equation}

Interestingly, the above implies that for $k_{\rm vir }R\ll 1$, a single resonance can have a major effect. (We do note that within the perfectly uniform spherical Earth approximation, when $k_{\rm vir}R\ll 1$, a fine-tuning of $f_a$ to other parameters with an accuracy of $(\kvir R)^2$ is required for a resonance to exist within the range of axion momenta spanned by the momentum integral.) Conversely, if we take $k_{\rm vir} R\gg 1$, no single resonance can contribute an amount $\gg 1$, unless $\dm R>(\kvir R)^3$. To see this, we note that $(1+2\ell)$ is at most of the order of $2\kdres R$ (since otherwise there are no resonances). Therefore, the result is at most of the order of $\sim e^{-\kres^2/\kvir^2}\kdres^3/(\kvir^3 R^2{\dm}^2)$. For any $\kres$, the expression $e^{-\kres^2/\kvir^2}\kdres^3=e^{-\kres^2/\kvir^2}(\kres^2+\dm^2)^{3/2}$ is at most of the order $(\kvir^2+\dm^2)^{3/2}$, so we can use that as a rough upper bound. Therefore, we remain with an expression that is at most of the order of $(\kvir^2+\dm^2)^{3/2}/(\kvir^3 R^2\dm^2)$. When $\dm\lesssim \kvir$, that expression is roughly $1/(\dm R)^2$, which is a small number. 
Conversely, when $\dm \gtrsim \kvir $, it is roughly $\dm/(\kvir ^3R^2)$. 
This leads to our original conclusion. 

Finally, it may be tempting to estimate the total number of resonances. However, the narrow width of the resonances, and their extreme sharp dependence on $\dm R$, implies our calculations are not robust against minuscule deviations from the uniform, static, spherically symmetric Earth approximation. Hence, despite being able to make some estimates using the asymptotic behaviors of spherical Bessel functions~\cite{}, we do not provide it here. We only note that for $\kvir R\ll 1$, the distance between two resonances is on average far larger than the range of $\kd$ scanned by the virial integral. Hence, one does not expect to see resonances there, unless $\dm R$ is fine tuned.

\section{Near-Threshold Bound States and Sommerfeld Enhancement}\label{app:sommer}

For all modes, we expect a resonant enhancement when there's a significant overlap with a bound state of the potential. When the momentum is zero, an overlap with the threshold bound state can be large. However, for a given ${\dm}^2$, not all $\ell$-modes might have a corresponding bound state at all. In fact, the first bound state appears when ${\dm} \, R \geq \pi/2 $, and will correspond to the $\ell=0$ mode at the threshold. 
Since higher-order $\ell$ terms have an additional centrifugal potential compared to the $\ell=0$ case, a deeper potential well is needed to yield a bound state for them.

The bound states are described by expressions similar to the ones encountered for the scattered wave (cp. Eq.~\eqref{eq:scatt}). They have an energy $\omega$ which satisfies $m_a^2-\dm^2\leq\omega^2\leq m_a^2$, hence the momentum outside is imaginary, $k_{\rm bound}^2=(i\kappa)^2$ for real $\kappa$, while $\kd^2{}_{\rm bound}=\dm^2-\kappa^2\geq 0$. Inside Earth, we still have
\begin{align}
    a^{\rm in}_{\omega,\ell}\left(r,\theta\right) =   b_{\ell}i^\ell\leri{2\ell+1}P_\ell\leri{\cos\theta} j_\ell\leri{\kd{}_{\rm bound} r}\,,
\end{align}
though for bound states, the full solution does not have a sum over $\ell$ (in fact each $\ell$ bound state has a $2\ell+1$ degeneracy, with $P_\ell(\cos(\theta))$ replaced by the spherical harmonics $Y_{\ell}^m(\theta,\phi)$).
Outside, the field should decay exponentially, which means that, it should be proportional to a spherical Hankel function of the first or second kind, according to whether $\kappa$ is positive or negative, respectively. Choosing the former, we therefore find
\begin{align}
    a^{\rm out}_{\omega,\ell}\left(r,\theta\right) = c_{\ell}i^\ell\leri{2\ell+1}P_\ell\leri{\cos\theta} h_\ell\leri{i\kappa r}\,.
\end{align}
Matching the field and its first derivative at the surface, one finds $\frac{b_\ell}{c_\ell}=\frac{h_\ell(i\kappa R)}{j_\ell(\kd{}_{\rm bound} R)}$ and the bound state condition for a generic $\ell$,
\begin{align}
    \sqrt{{\dm}^2-\kappa^2}\frac{j'_{\ell}\leri{R\sqrt{{\dm}^2-\kappa^2}}}{j_{\ell}\leri{R\sqrt{{\dm}^2-\kappa^2}}} = i\kappa \frac{h'_{\ell}\leri{iR\kappa}}{h_{\ell}\leri{iR\kappa}} \ .
    ~\label{eq:bound_condition}
\end{align}
Focusing on near-threshold bound states, which have energy $\approx m_a$ and $\kappa R\ll1$, we obtain 
$i\kappa \frac{h'_{\ell}\leri{iR\kappa}}{h_{\ell}\leri{iR\kappa}}\rightarrow -\frac{\leri{\ell+1}}{R}$\,, and the bound state exists whenever $\dm R \frac{j'_{\ell}\leri{\dm R}}{j_{\ell}\leri{\dm R}} = -\leri{\ell+1}$\,, which is only possible when\footnote{For $\ell>0$ and $f=j,y$ or $h$, we recall that $f'_{\ell}\leri{z} = f_{\ell-1}\leri{z}-\frac{\leri{\ell+1}f_\ell\leri{z}}{z}$. Using $f_{\ell+1}\leri{z}+f_{\ell-1}\leri{z}=\frac{2\ell+1}{z}f_\ell(z)$, we also get $f'_{\ell}\leri{z} = \frac{lf_\ell(z)}{z}-f_{\ell+1}\leri{z}$, which is used later.} $j_{\ell-1}\leri{\dm R} =0$.

We can rearrange our writing of the solution which describes the unbound mode, to highlight the relation of its deformation to the bound states. Defining
\begin{align}
    {\Dl}(x,\xd) \equiv \frac{h'_\ell\leri{x}}{h_{\ell}\leri{x}}-\frac{{\xd}}{x}\frac{j'_\ell\leri{{\xd}}}{j_\ell\leri{{\xd}}}\,,
\end{align}
with $x=k R\,, {\xd} = q R$\,, the bound state condition becomes ${\Dl}(i\kappa R,\kd{}_{\rm bound}R)=0$, and we find that $Q_\ell = ikh_\ell\leri{x}{\Dl}(x,\xd) j_\ell\leri{{\xd}}$\,. Therefore, 
\begin{align}
a_{\omega,\ell}\left(R,\theta\right) &= -\frac{i^{\ell} (2 \ell+1)P_\ell\leri{\cos\theta} e^{-i t q}}{ix^2 h_\ell\leri{x}{\Dl}(x,\xd)}   \,,\\
\frac{\vec{\grad}{a}_{\omega,\ell}\left(R,\theta\right)}{{a}_{\omega,\ell}\left(R,\theta\right)} &= \frac{q j'_\ell\leri{{\xd}}}{j_\ell\leri{{\xd}}}\,.
\end{align}

For the $\ell=0$ mode, the solution for the field is found in Eq.~\eqref{eq:a0}, and we find that when $k\rightarrow0$\,, this mode is enhanced as $1/\leri{kR}$ at $q R = n\pi+\pi/2$. This is consistent with the bound state condition at threshold for $\ell=0$ which reads $j_{-1}(\kd R)=-y_0(\kd R)=\frac{\cos(\kd R)}{\kd R}=0$. Maintaining a finite real momentum, we see that, even if a resonance appears, there is a deviation from the bound state corresponding to $1/\leri{\sqrt{k^2+\kappa^2} R}$\,.

While in principle such resonances can occur for any $\ell$, one should also take into account the overlap between the initial state and the bound state. When solving the naive $k=0$ problem, assuming a spherical symmetry of the solution coincided with requiring that the field is constant at $r\rightarrow \infty$\,, which corresponds to the $\ell=0$ mode at the $k\rightarrow 0$ limit. Higher $\ell$-modes may still resonate and contribute even for $k=0$, despite not being spherically symmetric (they correspond to higher excited states, with well-defined spherical quantum numbers).  However, we expect and find these modes to be suppressed in our $k=0$ problem, because they have very little overlap with the asymptotic state that resulted from a free wave at $r\rightarrow\infty$ when taking the strict $k=0$ limit. 

At small $kR$, we find
 \begin{align}
\text{Re}\leri{Q_\ell} &\rightarrow -\frac{2^{\ell-1} \Gamma \left(\ell+\frac{1}{2}\right) k\[2{\xd}j_{\ell-1}({\xd})-x^2j_\ell({\xd})\]}{\sqrt{\pi } x^{2+\ell}}\\
\text{Im}\leri{Q_\ell}& \rightarrow  \frac{2^{-\ell-1}x^\ell {\kd} \sqrt{\pi } j_{\ell+1}({\kd} R)}{\Gamma \left(\ell+\frac{3}{2}\right)}\,,
\end{align}
For the $k\to 0$ resonances, i.e., $j_{\ell-1}\leri{q R} =0$\,, we obtain $\text{Re}\leri{Q_\ell}\propto k x^{-\ell}$\,, and $\text{Im}\leri{Q_\ell}\propto  q x^{\ell}$.  Therefore, we obtain for $\ell\geq1$ a scaling of 
\begin{align}
a_{\ell}\leri{R}\Big|_{j_{\ell-1}\leri{q R} =0}\sim \leri{kR}^{\ell-2}
\,.
\end{align}

We note that the analogy presented here also hints at a regulatory effect not accounted for. Much like in Sommerfeld enhancement, the finite lifetime of the bound states (all but the ground state) is likely to introduce some suppression to the height of the resonances.

\section{Perturbation Theory}\label{app:pert}

For small enough ${\dm} R$ (as we will later see, compared to some power of $kR$), we may take a perturbation theory approach to our solution, 

\beq
a=a_{\rm free}({\vec r})+({\dm} R) ^2a^{(1)}({\vec r})+...
\eeq

We focus on $r=R$, since that is where most experiments are placed. Without the asymptotic Bessel function formulas~\cite{besselcite}, calculating $a^{(1)}$ for $kR\gg \mathcal{O}(10)$ is highly non-trivial, since the sum over $\ell$ modes oscillates, and thus a highly numerically stable algorithm is required. However, clear trends can be observed even at $kR\sim\mathcal{O}(10)$, which we expect to persist. 

\begin{figure}[h]
    \centering\includegraphics[width=0.9\linewidth]{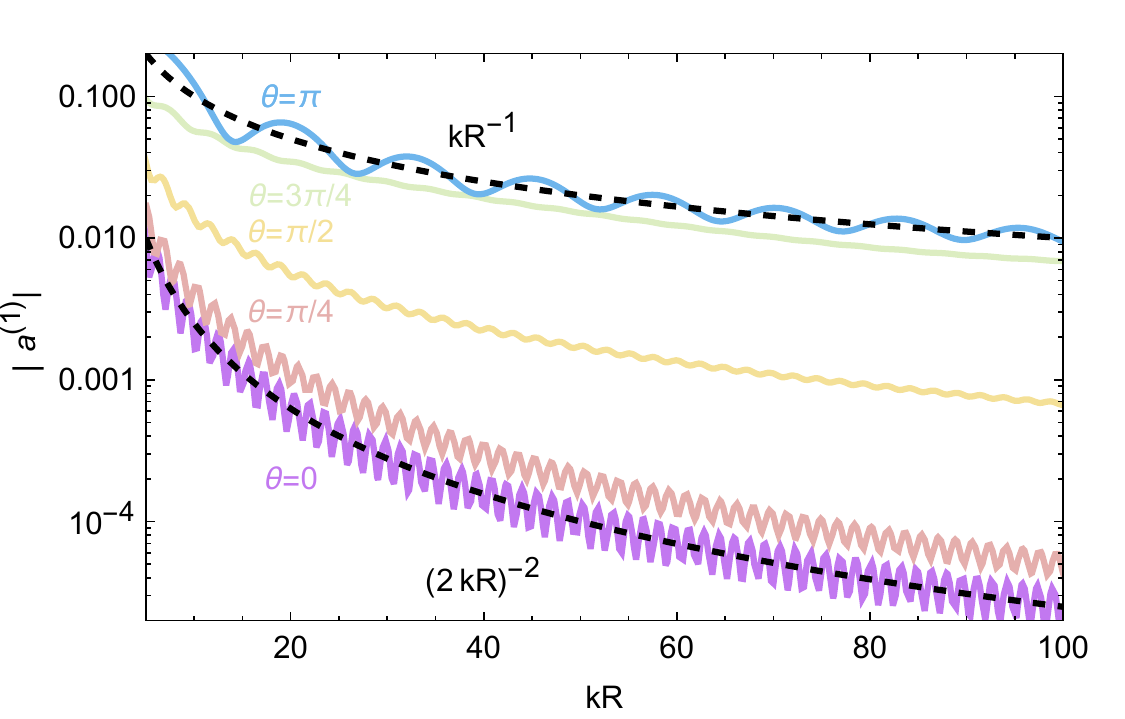}
    \caption{Doing perturbation theory for the field amplitude directly. A large part of the contribution is coming from a phase shift induced by the matter corrections. We examine different relative angles with $\cos\theta={\hat k}\cdot{\hat r}$. The black dashed lines are shown for illustration. As can be seen, for the antipodal point, the correction in the field amplitude (mostly from a phase shift), scales as $(\dm R)^2/(kR)$.}
    \label{fig:pert1}
\end{figure}

We begin by examining the magnitude of the correction to $a_{\omega}(R,\theta)$ for a few sampled $\theta$s in Figure~\ref{fig:pert1}. Ostensibly, for $\theta=\pi$ it seems major corrections should be present along the gray line of Figure~\ref{fig:parspace}. However, this is somewhat misleading. The naive expectation is that the solution inside Earth behaves like $e^{i\kd r}$, which changes the phase without changing the magnitude. Hence, it is reasonable that the phase correction is larger than the correction in $|a|^2$, discussed below.

\begin{figure}[h]
    \centering\includegraphics[width=0.9\linewidth]{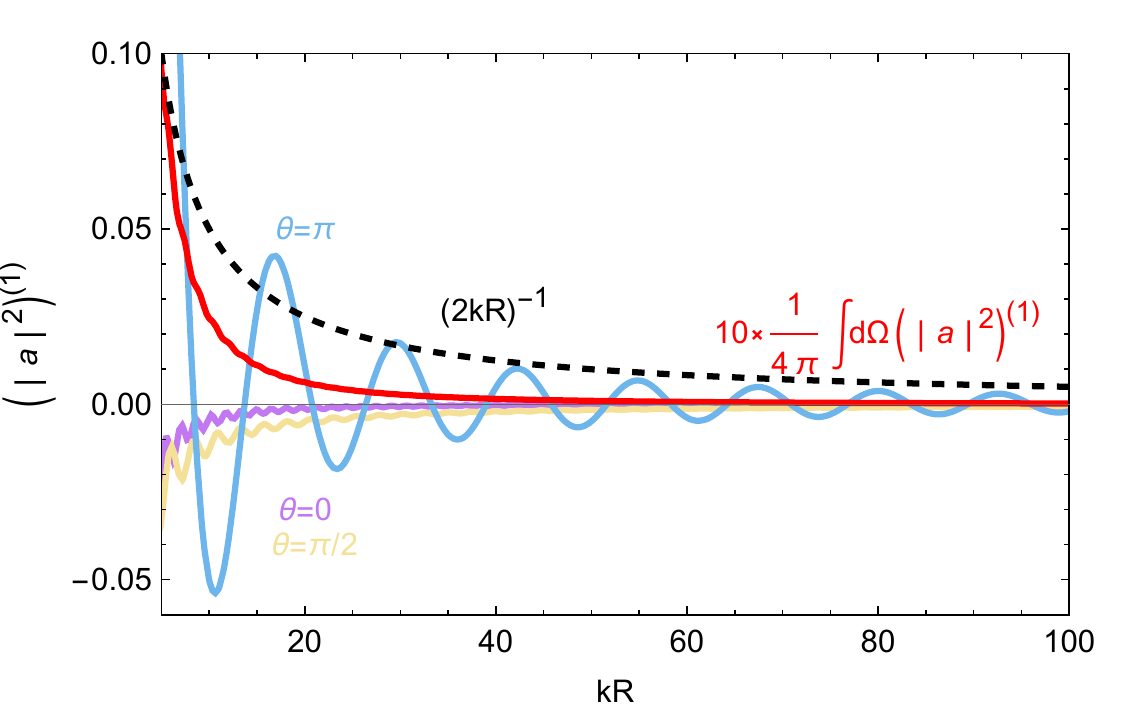}   
    \caption{The perturbation theory results for $|a|^2$. We examine different relative angles with $\cos\theta={\hat k}\cdot{\hat r}$, as well as the angularly averaged term. The latter roughly matches  $1/(2kR)^2$. The slowest reduction in amplitude is for $\theta=\pi$, and is faster than $(kR)^{-1}$, but slower than $(kR)^{-2}$, demonstrating the limitations of focusing on the angularly averaged quantities.}
    \label{fig:pert2} 
\end{figure}

Indeed, Figure~\ref{fig:pert2} shows the leading order correction (which is to be multiplied by $(\dm R)^2$) of the field magnitude, $|a|^2$. As can be seen, the change is milder, and at the antipodal point, it is mildly faster than $0.5/(kR)$. Conversely, averaging over the angle seems to highly suppress the correction, and matches $0.25/(kR)^2$, indicating that for $kR\gg \dm R$, no matter-effect are present in the averaged quantity. It should be noted that it is not obvious whether leading order perturbation theory captures high-$\ell$ resonances.

\begin{figure}[h]
    \centering
 \includegraphics[width=0.9\linewidth]{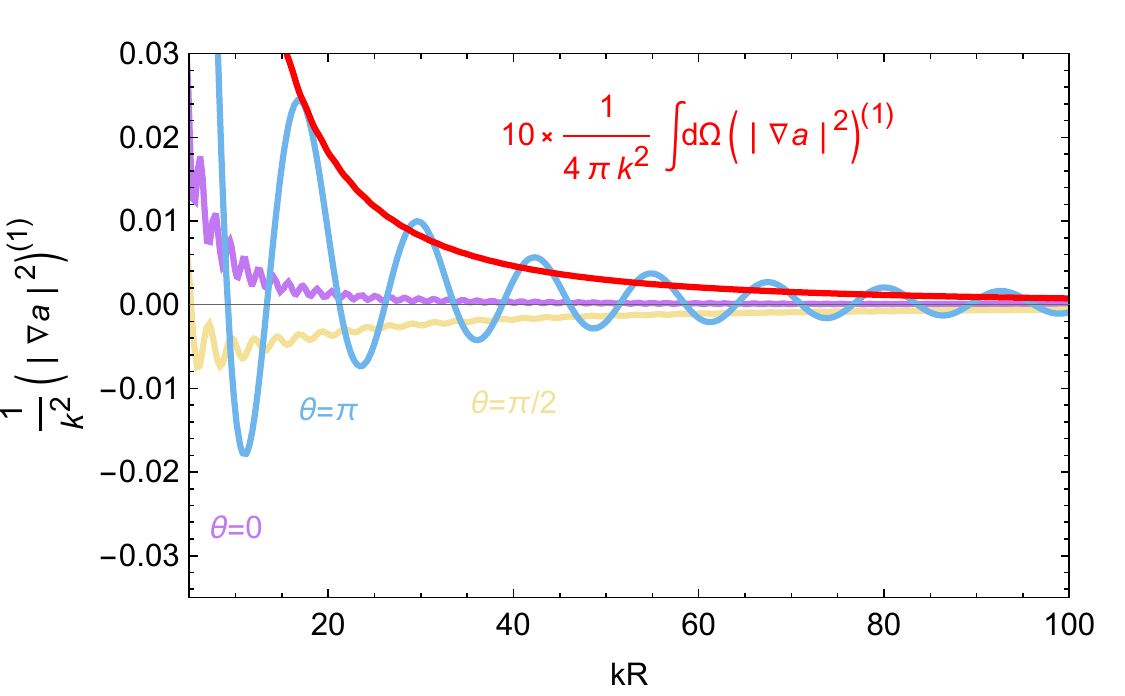}
    \caption{The perturbation theory results for $|\nabla a|^2$. We examine different relative angles with $\cos\theta={\hat k}\cdot{\hat r}$, as well as the spherically symmetrically virially averaged term. The spherical averaging matches roughly a constant (multiplied by 10 in the plot for comparison). }
    \label{fig:pert3}
\end{figure}

Finally, in Figure~\ref{fig:pert3}, we show the relative change in the field gradient compared to its zeroth order value. As in Figure~\ref{fig:pert2}, the angular averaging seems to mildly reduce the effect. However, here, the antipodal point's correction seems to be milder as well, scaling roughly in the vicinity of $\propto (kR)^{-1.5}$
.

Crucially, at least for $\int d\Omega |a|^2$, the correction is negligible when ${\dm}\ll k$. This shows the limitations of doing the spherically symmetric averaging. 

\section{Detectability of Earth-Affected Axions Coupled to Photons}\label{app:Vpt2}

Following the discussion in Section~\ref{sec:ExpConseq}, we discuss here some additional aspects of the detectablity of photon-coupled axions in the Earth-affected regions.

In the presence of the axion-photon interaction, Maxwell's equations are modified and read
\begin{align}
    \nabla\cdot \boldsymbol{E} &= \rho - g_{a\gamma\gamma}\boldsymbol{B} \cdot \grad a \ , \\ 
    \nabla \cdot \vect{B} &= 0 \ , \\
    \nabla \times \vect{E} + \partial_t \vect{B} &= 0 \ , \\
    \nabla \times \vect{B} - \partial_t \vect{E} &= \vect{j} - g_{a\gamma\gamma} \left(\vect{E} \times \grad a - \partial_t a \,\vect{B} \right) \ .
\end{align}
It is useful to write these as second-order differential equations. Furthermore, we can begin with the zeroth order equations, 
\begin{align}
    (\nabla^2 -\partial_t^2) \vect{E}_0 &= \grad \rho + \partial_t \vect{j} \ , \\
    (\nabla^2 -\partial_t^2) \vect{B}_0 &= -\nabla \times \vect{j} \ ,
\end{align}
creating the background  \ac{EM} fields, and only then solve the equations to $\mathcal{O}(g_{a\gamma\gamma})$,
\begin{align}
    \label{eq:E1eomAPP}
    (\nabla^2 -\partial_t^2) \vect{E}_1 &= -g_{a\gamma\gamma} \left( \grad\left(\vect{B}_0 \cdot \grad a \right) +\partial_t\left(\vect{E}_0 \times \grad a - \partial_t a \,\vect{B}_0 \right)\right) \ , \\
    (\nabla^2 -\partial_t^2) \vect{B}_1 &= g_{a\gamma\gamma}\nabla \times \left(\vect{E}_0 \times \grad a - \partial_t a \,\vect{B}_0 \right)  \ ,
    \label{eq:B1eomAPP}
\end{align}
 which are responsible for the signal. In the main text, we work with this set of equations.

Experiments are sensitive to an axion-induced signal power that scales as $P_{\rm sig} \sim  (\omega_{\rm sig}/Q)|\vect{B}_{1}|^2 V$. The quantity $\omega_{\rm sig}$ is the axion frequency $\omega$ for DC MQS experiments, and is $\omega_n \gg \omega$ for heterodyne experiments. Neglecting the effect of Earth, this would give $P_{\rm sig} \sim (Q \omega_{\rm sig}/\omega^2) (B_0 g^0_{a\gamma\gamma})^2 \rho_0 V$. Accounting for the effect of Earth and assuming equal sensitivity to $E$ and $B$ fields, we find
\begin{align}
    P_{\rm sig}^\oplus \sim \frac{Q\, \omega_{\rm sig}}{\omega^2} (B_0 g_{a\gamma\gamma}^\oplus)^2 \rho_0 V \left(\min \left[1, \frac{f_a}{(f_a)_\oplus^c} \right]^2 + v^2 \left(1 + \text{max}\left[ 1, \frac{1}{k R_\oplus}\right]\,\min \left[1, \left(\frac{(f_a)^c_\oplus}{f_a}\right)^2 \right] \right)^2 \right) \ .
    \label{eq:PsigEarth}
\end{align}
For DC experiments, the first term above arises from the $B$-field signal, while the second arises from the $E$-field signal.
For heterodyne experiments, both terms appear in $E$ and $B$, but the exact meaning of $Q$ and the scaling with $\omega$ should be computed more carefully. We refer the interested reader to~\cite{Berlin:2019ahk} for more details.
We observe that the first term in Eq.~\eqref{eq:PsigEarth} contributes a signal power that is independent of $f_a$ if $f_a \lesssim (f_a)^c_\oplus$. This conclusion is robust if $k R_\oplus \lesssim 1$, and may hold for $k R_\oplus \gtrsim 1$ also. Therefore, in the blue-green region of Fig.~\ref{fig:parspace}, it would be the second term, arising due to $\grad a(R_\oplus)$ that would dominate the signal. 
This signal is only detectable for DC MQS experiments if their readout is sensitive to the electric field signal. For heterodyne experiments, a non-zero overlap between the signal mode and $\hat{E}_0\times \hat{r}_\oplus$ is required.

\begin{figure}[t]
\centering
\includegraphics[width = 0.48\linewidth]{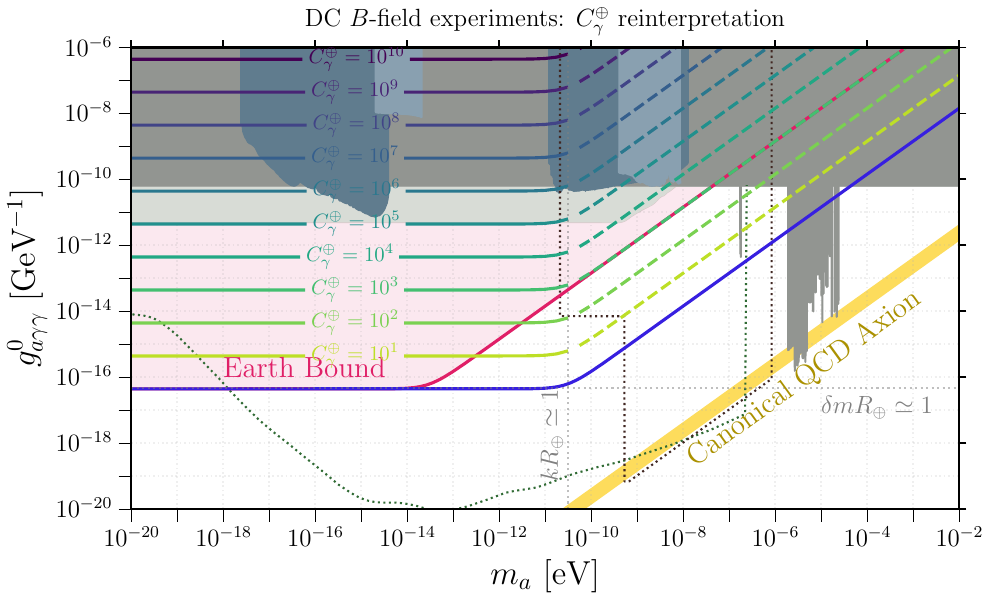}~
\includegraphics[width = 0.48\linewidth]{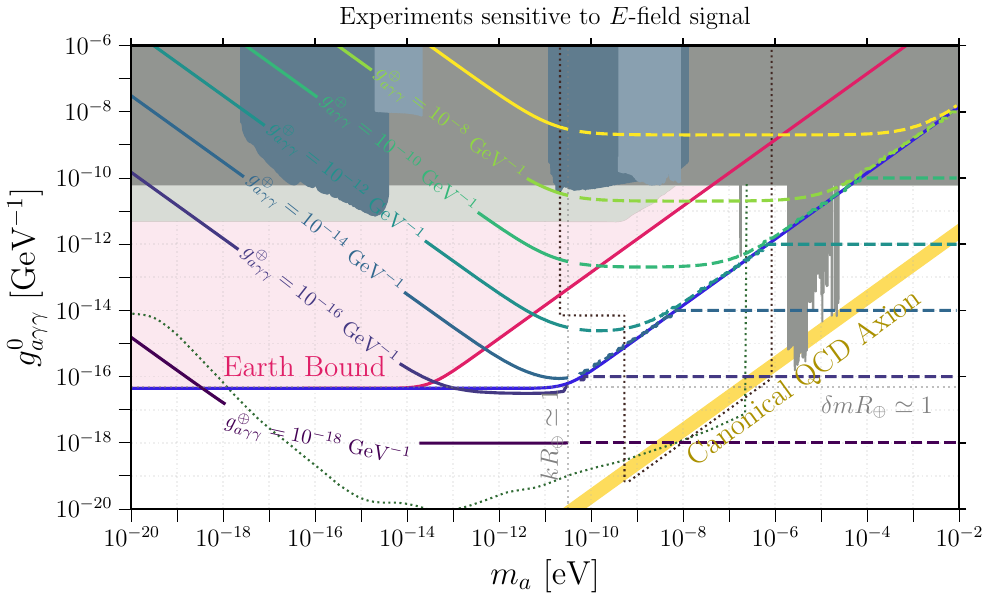}
\caption{\textbf{Left panel:} We show contours of $C_\gamma^\oplus$ that are being excluded by a corresponding $g_{a\gamma\gamma}^0$ limit set by an experiment. The Earth-affected axions impart a signal that is independent of $f_a$, and depends only on $(f_a)^c_\oplus$ and $C_\gamma^\oplus$. Therefore, to have a signal with a fixed $g_{a\gamma\gamma}^0$ requires a corresponding $C_\gamma^\oplus$.\\
\textbf{Right panel:}
If we assume an experiment is sensitive to both the $\partial_t a$-induced magnetic and the $\vect{\hat{r}}\cdot \grad a(R_\oplus)$-induced electric fields, then an experiment sensitive to a given $g_{a\gamma\gamma}^0$ is actually sensitive to $g_{a\gamma\gamma}^\oplus$ shown by the corresponding contour, assuming $C_\gamma = 1$.
\\
Shown in both panels are existing constraints set under the assumption $a(R_\oplus) = a_0$~\cite{PhysRevLett.104.041301,ADMX:2018gho,Ouellet:2018beu,Crisosto:2019fcj,Lee:2020cfj,Jeong:2020cwz,CAPP:2020utb,Gramolin:2020ict,Salemi:2021gck,Arza:2021ekq,ADMX:2021nhd,Sulai:2023zqw,ADMX:2024xbv,CAPP:2024dtx,Friel:2024shg}, as well as projections of DMRadio~\cite{DMRadio:2022jfv,DMRadio:2022pkf,DMRadio:2023igr} and a Heterodyne SRF experiment~\cite{Berlin:2019ahk,Berlin:2020vrk,Giaccone:2022pke}.
}
\label{fig:PhotonExperimentsReinterp}
\end{figure}

If we take the first term of Eq.~\eqref{eq:PsigEarth} only, we see that
the experimentally-probed combination of $g_{a\gamma\gamma} a(R_\oplus)$ becomes independent of $f_a$ in the Earth-affected region (blue in Fig.~\ref{fig:parspace}, so that any existing constraints on axion \ac{DM} should be re-interpreted as a constraining on $C_\gamma$. In the left panel of Fig.~\ref{fig:PhotonExperimentsReinterp} we show how such a re-interpretation could look. For a given $g_{a\gamma\gamma}^0$, i.e., the coupling that is constrained experimentally assuming axion \ac{DM} that is unaffected by Earth, we show contours of the corresponding $C_\gamma^\oplus$ if $a(R_\oplus)$ is affected. We see that in the affected region (above the blue contour) that is not excluded by the Earth Bound (red shaded region), existing DC magnetic field experiments are constraining $C_\gamma \lesssim 10^3$, with $f_a = (f_a)_\oplus^c$. This includes almost all the parameter space that DMRadio-50L anticipates probing~\cite{Rapidis:2022gti}, and the region currently excluded by ADMX-SLIC~\cite{Crisosto:2019fcj}. It also covers a substantial portion of the parameter space that a heterodyne SRF experiment could probe in the most optimistic projections of Refs.~\cite{Berlin:2019ahk,Berlin:2020vrk}.

This effect is mitigated if experiments are sensitive to the $\vect{\hat{r}}\cdot \grad a(R_\oplus)$-enhanced \emph{electric} field signal, i.e. the second term in Eq.~\eqref{eq:PsigEarth}. In that case, for small axion masses the Earth-affected sensitivity can exceed the na\"ive vacuum axion \ac{DM} sensitivity. This is shown in the right panel of Fig.~\ref{fig:PhotonExperimentsReinterp}. We show contours of $g_{a\gamma\gamma}^\oplus$, the coupling constant probed on Earth, corresponding to the equivalent naive combination $g_{a\gamma\gamma}^0 a_0$ set by the $y$-axis and the assumptions of axion \ac{DM} and $C_\gamma = 1$. We see that a contour of fixed $g_{a\gamma\gamma}^\oplus$ rises above the corresponding $g_{a\gamma\gamma}^0$ for sufficiently small axion masses. However, we remind the reader that the red-shaded region is excluded by the Earth bound, so that most of the region affected by this enhanced sensitivity is excluded unless the Earth bound can somehow be evaded. The exception to this is the $g_{a\gamma\gamma}^\oplus = 10^{-18}\,\text{GeV}^{-1}$ contour, which we see exhibits a mild enhancement for axion masses below $m_a \lesssim 10^{-17}\,\text{eV}$. Unfortunately, this region is below the boundary of the most optimistic SRF estimates. We caution that we have set $C_\gamma = 1$ in the right panel of Fig.~\ref{fig:PhotonExperimentsReinterp}. Since the Earth bound (and the region above the blue contour) depend on $f_a$ and not $g_{a\gamma\gamma}$, the interpretation of these regions should be performed with care.

\bibliographystyle{JHEP}
\bibliography{biblioArxiv}

\end{document}